\documentclass[review,12pt]{elsarticle}

\setcitestyle{square}
\usepackage{amssymb}
\usepackage{amsthm}
\usepackage{amsmath}
\usepackage{breqn}
\usepackage{mathrsfs}
\usepackage{epsfig}
\usepackage{upgreek} 
\usepackage{graphicx}
\usepackage{epstopdf}
\usepackage{float}
\usepackage{caption}
\usepackage{subcaption}
\usepackage{bm}
\usepackage{bbm}
\usepackage{mathrsfs}
\usepackage{mathtools}
\usepackage{cleveref}
\usepackage{soul}
\usepackage{accents}
\usepackage{xcolor}
\usepackage{courier} 
\usepackage{listings} 
\usepackage[margin=2cm]{geometry}
\usepackage{tabu} 
\usepackage{longtable}
\usepackage[T1]{fontenc}
\usepackage{changepage} 
\biboptions{sort&compress} 

\journal{Engineering Fracture Mechanics}

\makeatletter
\def\@author#1{\g@addto@macro\elsauthors{\normalsize%
    \def\baselinestretch{1}%
    \upshape\authorsep#1\unskip\textsuperscript{%
      \ifx\@fnmark\@empty\else\unskip\sep\@fnmark\let\sep=,\fi
      \ifx\@corref\@empty\else\unskip\sep\@corref\let\sep=,\fi
      }%
    \def\authorsep{\unskip,\space}%
    \global\let\@fnmark\@empty
    \global\let\@corref\@empty  
    \global\let\sep\@empty}%
    \@eadauthor={#1}
}
\makeatother

\begin{document}

\begin{frontmatter}



\title{A stress-based poro-damage phase field model for hydrofracturing of creeping glaciers and ice shelves}


\author[IC]{Theo Clayton}

\author[VU]{Ravindra Duddu}

\author[G,ES]{Martin Siegert}

\author[IC]{Emilio Mart\'{\i}nez-Pa\~neda\corref{cor1}}
\ead{e.martinez-paneda@imperial.ac.uk}

\address[IC]{Department of Civil and Environmental Engineering, Imperial College London, London SW7 2AZ, UK}

\address[VU]{Department of Civil and Environmental Engineering, Vanderbilt University, Nashville, TN 37235, USA}

\address[G]{Grantham Institute, Imperial College London, London, SW7 2AZ, UK}
\address[ES]{Department of Earth Science and Engineering, Imperial College London, London, SW7 2AZ, UK}

\cortext[cor1]{Corresponding author.}

\begin{abstract}
There is a need for computational models capable of predicting meltwater-assisted crevasse growth in glacial ice. Mass loss from glaciers and ice sheets is the largest contributor to sea-level rise and iceberg calving due to hydrofracture is one of the most prominent yet less understood glacial mass loss processes. To overcome the limitations of empirical and analytical approaches, we here propose a new phase field-based computational framework to simulate crevasse growth in both grounded ice sheets and floating ice shelves. The model incorporates the three elements needed to mechanistically simulate hydrofracture of surface and basal crevasses: (i) a constitutive description incorporating the non-linear viscous rheology of ice, (ii) a phase field formulation capable of capturing cracking phenomena of arbitrary complexity, such as 3D crevasse interaction, and (iii) a poro-damage representation to account for the role of meltwater pressure on crevasse growth. A stress-based phase field model is adopted to reduce the length-scale sensitivity, as needed to tackle the large scales of iceberg calving, and to adequately predict crevasse growth in tensile stress regions of incompressible solids. The potential of the computational framework presented is demonstrated by addressing a number of 2D and 3D case studies, involving single and multiple crevasses, and considering both grounded and floating conditions. The model results show a good agreement with analytical approaches when particularised to the idealised scenarios where these are relevant. More importantly, we demonstrate how the model can be used to provide the first computational predictions of crevasse interactions in floating ice shelves and 3D ice sheets, shedding new light into these phenomena. Also, the creep-assisted nucleation and growth of crevasses is simulated in a realistic geometry, corresponding to the Helheim glacier. The computational framework presented opens new horizons in the modelling of iceberg calving and, due to its ability to incorporate incompressible behaviour, can be readily incorporated into numerical ice sheet models for projecting sea-level rise.\\
\end{abstract}


\end{frontmatter}


\section{Introduction}
\label{Introduction}


Ice sheets are large masses of glacial ice that inundate the surrounding landscape in Greenland and Antarctica today, and many other regions during ice ages \cite{Siegert2001}. These act as enormous stores of freshwater - containing approximately 70\% of the planet's supply \cite{Siegert2020} - that assist in regulating a stable global climate, through maintaining global ocean-water levels and controlling surface temperatures by reflecting solar radiation due to its high albedo properties \cite{Siegert2005}. Ice sheets thin toward their margins, and if these are located in marine settings, they will form floating extensions known as ice shelves, which act to provide resistive buttressing to downslope flow and reduce the flux of grounded ice to the ocean. However, increasing global temperatures as a result of carbon emissions has lead to higher rates of ablation than accumulation, resulting in ice shelf and ice sheet thinning in some key areas where ice-sheet instability may follow \cite{Mercer1978}. Surface and basal crevasses can form within ice sheets as a consequence of ongoing deformations within the ice. These are deep crack-like defects that can propagate in an unstable manner and lead to large-scale iceberg calving events, and in extreme cases the catastrophic break up of ice shelves. The frequency of these events has grown in recent decades, beginning with the disintegration of Larsen A (1995) \cite{Rott1996RapidAntarctica} and Larsen B (2002)  \cite{Domack2005} ice shelves, and more recently significant surface melting and iceberg calving on Larsen C (2017) \cite{Mitcham2022TheShelf}, Pine Island and Thwaites (2018-2020) \cite{Lhermitte2020DamageEmbayment}, and Conger (2022) ice shelves. Fracture within ice shelves can result in a loss of resistance to down slope glacial flow, leading to ice-sheet thinning, additional flotation of grounded ice and, thus, potentially irreversible grounding line retreats \cite{Doake1998}.\\ 

\begin{figure} [H]
    \centering
    \includegraphics[width=0.8\textwidth]{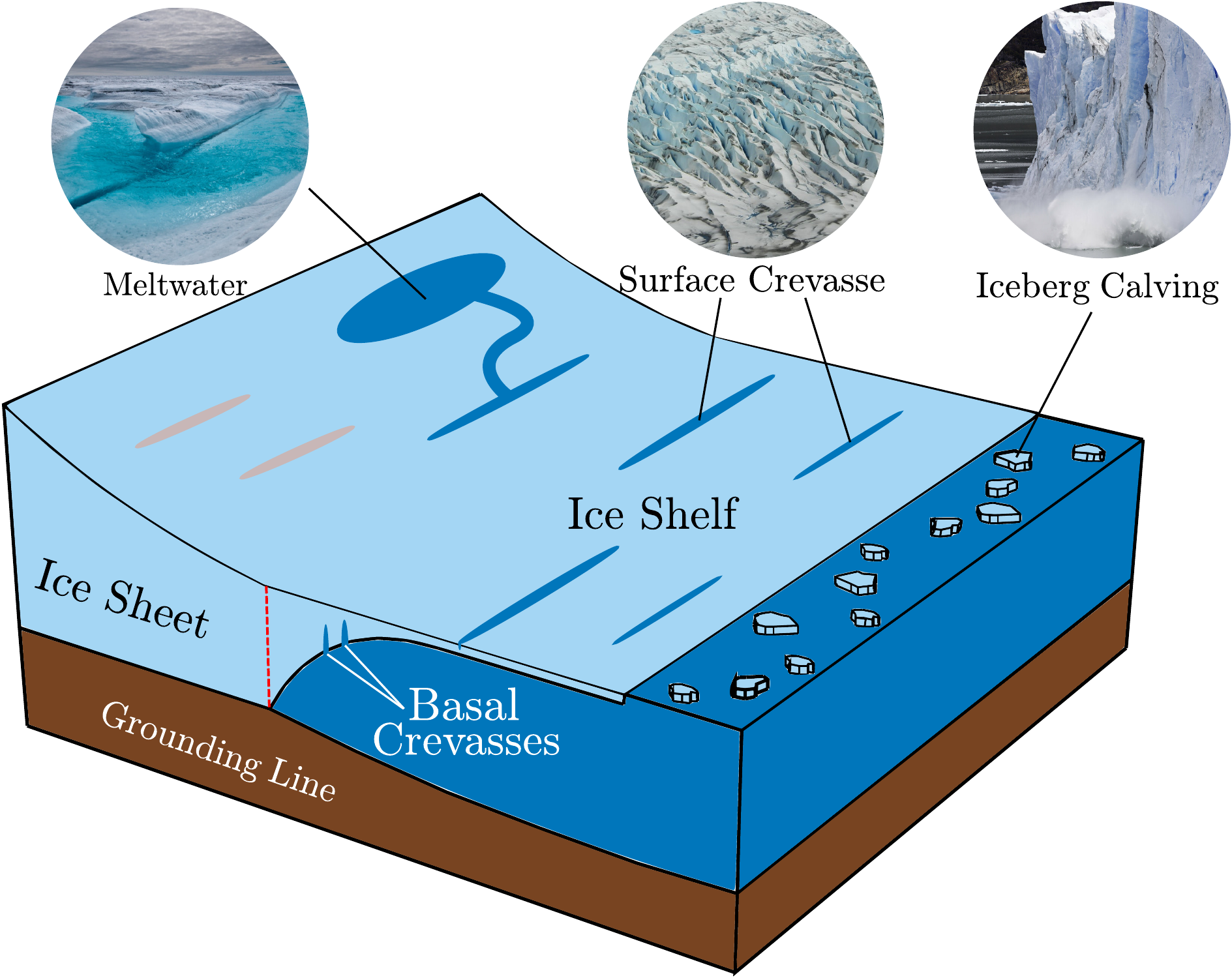}
    \caption{Illustration of a grounded ice sheet and a floating ice shelf, containing both surface and basal crevasses, and with calving events occurring at the terminus.}
    \label{fig:3D_Diagram}
\end{figure}

Deposition of grounded glacial ice into the ocean is one of the leading contributors to sea level rise \cite{Frederikse2020The1900}, having direct implications within this Century on low-lying coastal regions through flooding, increased extreme environmental events, degradation of farmland and loss of habitat, among others. A key driving factor for their stability is the production of surface meltwater as a result of elevated surface temperatures \cite{Scambos2004GlacierAntarctica}. When ice shelves and glaciers melt, meltwater flows down-slope into surface crevasses, causing additional tensile stresses to form within the crevasse. This leads to crevasse instability, and with sufficient meltwater, the crevasse can propagate through the full thickness of the ice column. This process is generally referred to in the glaciological literature as hydrofracture \cite{Scambos2009IceBreak-ups}. A recent study by Lai et al. \cite{Lai2020} found that approximately 60 $\pm$\ 10 $\%$ of Antarctic ice shelves provide significant buttressing to downslope flow and are vulnerable to meltwater driven hydrofracture, highlighting the significance of studying the formation and propagation of crevasses in glaciers. An illustration of a grounded ice sheet, transitioning to a crevassed floating ice shelf is shown in Fig. \ref{fig:3D_Diagram}.

Ice sheet fracture and crevasse propagation have been mainly modelled previously using analytical methods. The estimation of crevasse penetration depths in an idealised glacier was first described by Nye in 1955 \cite{Nye1955CommentsCrevasses} based on the so-called 'zero stress' model. Nye assumed that ice has no tensile resistance to fracture and that a crevasse will stabilise at a depth where the longitudinal tensile stress is balanced by the lithostatic compressive stress \cite{Nye1957}. This was later extended by Benn et al. to include the presence of meltwater within crevasses \cite{Benn2007}. Linear elastic fracture mechanics models were introduced to provide a more accurate prediction of the depth of an isolated crevasse \cite{VanDerVeen1998, VanDerVeen1998a}. By exploiting the principle of superposition, stress intensity factors can be calculated by integrating over the crevasse depth for the normal tensile stress, the lithostatic compressive stress, and the meltwater pressure. In order for crevasses to stabilise, the net stress intensity factor $K_{\text{net}}$ must be equal to the material's fracture toughness $K_c$. However, these analytical approaches have well-known limitations, such as (i) idealised scenarios and boundary conditions are assumed; (ii) creep effects, resulting from the continual movement of glaciers under their own weight, are neglected; and (iii) crevasse interaction cannot be captured.\\

Recently, computational methods have been used to predict crevasse growth and iceberg calving events. Local and non-local continuum damage mechanics formulations have been presented to predict ice sheet fracture \cite{Pralong2005DynamicCalving,Duddu2012,Mobasher2016,Duddu2020}. These works have overcome some of the limitations intrinsic to analytical approaches, but often at the cost of using empirical parameters. Variational phase field fracture models offer an alternative approach, enabling the simulation of realistic conditions (3D geometries, multiple interacting crevasses, \textit{etc}.) and providing a connection to fracture mechanics theory. Phase field fracture models have gained remarkable popularity in recent years due to their ability to predict complex cracking phenomena including crack bifurcation, coalescence and nucleation from arbitrary sites \cite{Bourdin2008,TAFM2020,Molnar2022}. New phase field-based formulations have been presented for dynamic fracture \cite{Borden2012,Mandal2020b}, ductile damage \cite{Shishvan2021a,Aldakheel2021a}, environmentally assisted cracking \cite{JMPS2020,Wu2020b}, fatigue crack growth \cite{Carrara2020,CMAME2022}, hydraulic fracture \cite{Xia2017,Heider2021}, and battery degradation \cite{Xue2022,Boyce2022}; among other (see Refs. \cite{Wu2020,PTRSA2021} for an overview). In this work, we aim at extending the success of phase field fracture models to the area of glacier crevassing and iceberg calving. To this end, a new phase field formulation is presented capable of capturing the creep behaviour of glacial ice and the role of fluid pressure in driving crevasse growth. Also, for the first time, crevasse interaction is predicted in both 2D and 3D. Very recently, Sun \textit{et al.} \cite{Sun2021} used a phase field approach to predict hydrofracture in 2D linear elastic glaciers, assuming compressible behaviour and disregarding creep effects. Unlike them, we base our framework on a stress-based phase field fracture formulation, which offers several advantages in the context of hydrofracturing of glacier crevasses. First, strain energy-based approaches are unsuited for incompressible rheologies. This is not only important due to the incompressible nature of glacial ice, but also because it hinders its integration into large-scale computational models for ice sheet evolution and sea level rise, which assume incompressible flow (see, e.g., the Community Ice Sheet Model (CISM) \cite{Lipscomb2019}). Second, ice-sheet fracture is driven by tensile stresses and not strains, with crevasses propagating solely in regions where the net longitudinal stress is positive \cite{Smith1976}. This is naturally accounted for in a stress-based phase field model, while requiring a particular \textit{ad hoc} split in strain energy-based formulations \cite{Sun2021,Lo2019}. Third, a phase field length-scale insensitive driving force can be defined, enabling the use of coarser meshes, a key enabler given the large scales involved. These advantages provide further motivation for this work, presenting the first stress-based phase field computational framework for hydrofracturing of creeping glaciers and ice shelves.\\ 

The rest of the paper is outlined as follows. The theoretical and computational framework presented is described in Section \ref{Sec:Theory}. The model is then used in Section \ref{Sec:Results} to predict hydrofracturing in case studies of particular interest. First, the propagation of single crevasses in grounded ice considering both linear and non-linear rheologies is investigated. A parametric study is conducted to assess the role of relevant material parameters, seawater level and meltwater depth. Second, we simulate the growth of a field of densely spaced crevasses in a grounded glacier, comparing against the predictions of Nye's zero stress model. Third, the growth of basal and surface crevasses (and their interaction) is for the first time simulated for a floating ice shelf, using appropriate Robin boundary conditions. Fourth, the combined creep-phase field fracture model is used to predict the nucleation and growth of crevasses in a realistic geometry, corresponding to the Helheim glacier. Finally, we provide the first 3D analysis of crevasse propagation in ice sheets. Concluding remarks end the manuscript in Section \ref{Sec:Concluding remarks}. 

\section{Numerical framework} 
\label{Sec:Theory}

\begin{figure}
    \centering
    \includegraphics[width=0.75\textwidth]{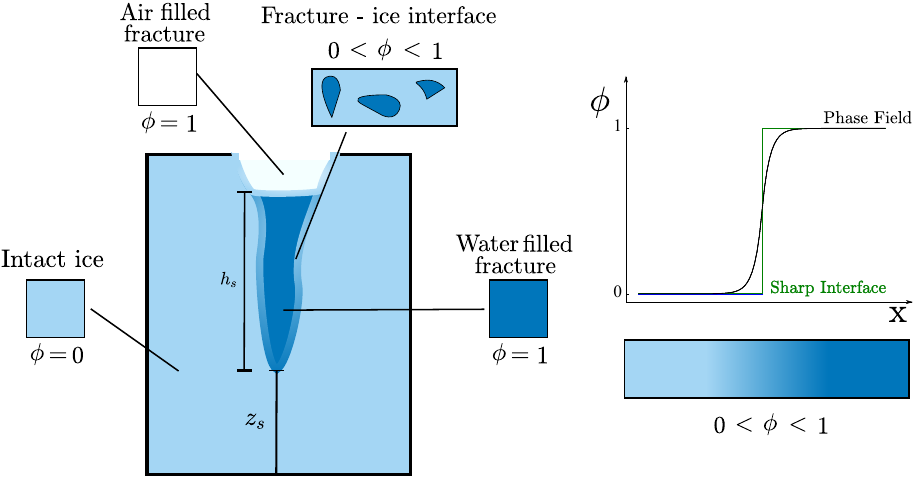}
    \caption{Schematic diagram of a meltwater filled crevasse in glacial ice, illustrating the intact phase ($\phi=0$), fully cracked phase ($\phi=1$) and transition phase ($0<\phi<1$). In the damaged and transition phases, there is a hydrostatic pressure contribution to damage arising from the meltwater. Relevant to the poro-damage part of the model, $h_s$ denotes the meltwater depth, and $z_s$ is the distance between the glacier base and the bottom of the crevasse, with $z$ being the vertical height.}
    \label{fig:Phase_Field_Ice_Diagram}
\end{figure}

In this section, we present our computational framework, which encompasses the three elements that are needed to resolve the hydrofracture process taking place in ice sheets; namely, the viscoplastic behaviour of ice, the propagation of meltwater-filled crevasses, and the role of meltwater pressure on crevasse propagation. These are modelled by means of Glen's flow law \cite{Glen1955}, a stress-based phase field description of fracture \cite{Miehe2015a}, and a meltwater-ice poro-damage model \cite{Mobasher2016}, respectively. Fig. \ref{fig:Phase_Field_Ice_Diagram} illustrates upon a single crevasse the mechanistic and modelling assumptions of our framework. In the following, we present the kinematics of the problem (Section \ref{Sec:Kinematics}), formulate the energy functionals (Section \ref{Sec:PVW}), particularise the model upon suitable constitutive choices (Section \ref{Sec:ConstitutiveTheory}), and briefly describe the finite element implementation (Section \ref{Sec:FEM}). Throughout, the formulation refers to a body occupying an arbitrary domain $\Omega \subset {\rm I\!R}^n$ $(n \in[1,2,3])$, with an external boundary $\partial \Omega\subset {\rm I\!R}^{n-1}$ with outwards unit normal $\mathbf{n}$.

\subsection{Kinematics and general considerations}
\label{Sec:Kinematics}

The primary variables are the displacement field vector $\mathbf{u}$ and the damage phase field $\phi$. Restricting our attention to small strains and isothermal conditions, the strain tensor $\bm{\varepsilon}$ reads
\begin{equation}
    \bm{\varepsilon} = \frac{1}{2}\left(\nabla\mathbf{u}^T+\nabla\mathbf{u}\right) \, ,
\end{equation}

\noindent with the strain field itself being additively decomposed into its elastic and viscous parts, such that 
\begin{equation}
    \bm{\varepsilon} = \bm{\varepsilon}^e + \bm{\varepsilon}^v \, .
\end{equation}

The growth of meltwater-filled crevasses is described by means of a smooth continuous scalar \emph{phase field}, which takes a value of $\phi=0$ in intact ice and of $\phi=1$ in fully damaged regions (see Fig. \ref{fig:Phase_Field_Ice_Diagram}). The aim is to overcome the need to track discrete crack surfaces, which is a cumbersome task. Thus, the intact ice-crack interface is not explicitly modelled but instead smeared over a finite domain, replacing interfacial boundary conditions by a differential equation that describes the evolution of the phase field $\phi$. The smearing of the interface is controlled by a phase field length scale $\ell$. Accordingly, a discontinuous surface $\Gamma$ is regularised through the following crack surface functional \cite{Bourdin2000}:
\begin{equation}
\Gamma_\ell \left( \phi \right) = \int_\Omega \gamma_\ell \left( \phi, \nabla \phi \right) \, \text{d}V \, ,
\end{equation}

\noindent where $\gamma_\ell$ is the so-called crack surface density functional.

\subsection{Energy functionals}
\label{Sec:PVW}

A total potential energy can be defined by incorporating the contributions from the bulk strain energy density $\psi_s$, which itself accounts for both viscous ($\psi_s^v$) and elastic ($\psi_s^e$) contributions, and the regularised fracture energy $\psi_f$. Thus, considering the work done by external tractions $\mathbf{T}$ and body forces $\mathbf{b}$, the total potential energy of the solid can be expressed as,
\begin{equation}
     \Psi_{pot} \left( \mathbf{u}, \phi \right) = \int_{\Omega}\left[\psi_s \left(\bm{u}, \phi \right) + \psi_f \left( \phi, \nabla \phi \right) \right] \,\text{d}V  - \int_\Omega \mathbf{b} \cdot \mathbf{u} \, \text{d} V - \int_{\partial \Omega} \mathbf{T} \cdot \mathbf{u} \, \text{d}S \, .
\end{equation}

As discussed below, it is important to consider as well the kinetic energy of the body, which is given by
\begin{equation}
  \Psi_{kin} \left( \dot{\mathbf{u}} \right) =  \frac{1}{2} \int_\Omega \rho \dot{\mathbf{u}}\cdot \dot{\mathbf{u}} \, \text{d}V \, ,
\end{equation}

\noindent where $\rho$ is the mass density of the material and $\dot{\mathbf{u}}=\partial \mathbf{u} / \partial t$. The Lagrangian for the coupled deformation-fracture problem can then be formulated by combining the kinetic and total potential energies, such that
\begin{equation}
    L \left( \mathbf{u}, \dot{\mathbf{u}}, \phi \right) = \Psi_{kin} \left( \dot{\mathbf{u}} \right) - \Psi_{pot} \left( \mathbf{u}, \phi \right)  = \int_\Omega \left[ \frac{1}{2} \rho \dot{\bm{u}}\cdot \dot{\bm{u}} - \psi_s \left(\bm{u} , \phi \right) - \psi_f \left( \phi, \nabla \phi \right) + \mathbf{b} \cdot \mathbf{u} +   \mathbf{T} \cdot \mathbf{u} \right] \text{d}V \, .
\end{equation}

We shall now make constitutive assumptions and, building upon these, proceed to formulate the local force balances.

\subsection{Constitutive theory}
\label{Sec:ConstitutiveTheory}

We proceed to particularise our choices with the aim of providing a suitable framework for predicting ice-sheet hydrofracture. To this end, the bulk strain energy density of the solid is given in terms of its elastic and viscous counterparts as,
\begin{equation}\label{eq:BulkSED}
    \psi_s = g \left( \phi \right) \psi_s^e \left( \bm{\varepsilon}^e \right) + \psi_s^v \left(  \bm{\varepsilon}^v \right) = g \left( \phi \right) \left\{ \frac{1}{2} \lambda \left[ \text{tr} \left( \bm{\varepsilon}^e \right) \right]^2 + \mu \text{tr} \left(  \bm{\varepsilon}^e \cdot    \bm{\varepsilon}^e \right) \right\} + \int_0^t \left( \bm{\sigma}_0 : \dot{\bm{\varepsilon}}^v \right) \, \text{d} V \, ,
\end{equation}

\noindent where $\bm{\sigma}_0$ is the undamaged Cauchy stress tensor, $\lambda$ and $\mu$ are the Lamé parameters, and $g \left( \phi \right)$ is a phase field degradation function, to be defined. Then, the homogenised (damaged) stress tensor can be estimated as $\bm{\sigma}=\partial_{\bm{\varepsilon}^e} \psi_s$. As described below, the viscous behaviour of the solid is described by Glen's flow law \cite{Glen1955}, a commonly used choice for glacial ice.

\subsubsection{Creep behaviour of ice: Glen's flow law}
\label{Sec:GlenFlowLaw}

Glacial ice is a polycrystalline material undergoing a state of constant stress and operating close to its melting point. It is therefore prone to creep. Creep deformation is a well documented process within glaciers and was first studied by Glen in 1955 \cite{Glen1955}. Glen proposed a steady state creep law based on the Bingham-Norton/Maxwell model, by which the viscous strain rates are given as
\begin{equation}
   \Dot{ \boldsymbol{\varepsilon}}^v = A \left( \sigma_e\right)^{n-1} \bm{\sigma}_0' \, ,
\end{equation}

\noindent where $A$ is the creep coefficient, $\bm{\sigma}_0'=\bm{\sigma}_0 - \text{tr}(\bm{\sigma}_0) \bm{I}/3$ is the undamaged deviatoric stress tensor, $n$ is the creep exponent, and $\sigma_e$ is an equivalent stress measure defined as $\sigma_e = \sqrt{\frac{1}{2}\bm{\sigma}_0' : \bm{\sigma}_0'}$. The creep coefficient $A$ and the creep exponent $n$ are typically calibrated with experiments, with the former exhibiting the following Arrhenius dependency with temperature,
\begin{equation}
    A = A_{0} \exp{\frac{Q}{RT}} \, ,
\end{equation}

\noindent where $T$ is the absolute temperature, $Q$ is the activation energy, $R$ is the universal gas constant, and $A_{0}$ is the creep coefficient at a reference temperature $T_{0}$.





\subsubsection{A stress-based phase field fracture model}

The evolution of damage is driven by the phase field variable $\phi$. A length-scale insensitive, stress-based approach is adopted, inspired by the work by Miehe \textit{et al.} \cite{Miehe2015a}. This choice enables us to capture purely stress-driven fractures in incompressible solids using relatively coarse meshes; as required to model hydrofractures in creeping glaciers. Accordingly, the fracture energy density is formulated as,
\begin{equation}\label{eq:psi_f}
    \psi_f \left( \phi, \, \nabla \phi \right) = 2 \psi_{c} \left( \phi + \frac{\ell^2}{2} \left| \nabla \phi \right|^2 \right) \, .
\end{equation}

Unlike conventional phase field fracture models, Eq. (\ref{eq:psi_f}) shows that the present formulation introduces the phase field through a linear term. This naturally results in a damage threshold, below which $\phi=0$, preserving the elastic properties of uncracked regions. In (\ref{eq:psi_f}), $\psi_c$ is a fracture energy density, which in a stress-based approach is defined as a function of a critical fracture stress or material strength $\sigma_c$, such that \cite{Miehe2015a}:
\begin{equation}
    \psi_c=\frac{\sigma_c^2}{2E} \, .
\end{equation}

\noindent Here, $E$ denotes the material's Young's modulus. It remains to define the degradation function $g(\phi)$, which reduces the elastic stiffness of the solid - see Eq. (\ref{eq:BulkSED}). The choice of $g(\phi)$ must fulfill the following conditions,
\begin{equation}
g \left( 0 \right) =1 , \,\,\,\,\,\,\,\,\,\, g \left( 1 \right) =0 , \,\,\,\,\,\,\,\,\,\, g' \left( \phi \right) \leq 0 \,\,\, \text{for} \,\,\, 0 \leq \phi \leq 1 \, .
\end{equation}

\noindent Here, we choose to adopt the following quadratic degradation function
\begin{equation}
    g \left( \phi \right) = \left( 1 - \phi \right)^2 \, .
\end{equation}

Finally, the phase field evolution law is given by \cite{Miehe2015a},
\begin{equation}\label{eq:PhiForceBalance}
    \phi - \ell^2 \nabla^2 \phi = 2 \left( 1 - \phi \right) D_d \, .
\end{equation}

Where the left hand side is the geometric resistance and the right hand side corresponds to the driving force. Here, $D_d$ is the crack driving force state function, which is here defined based on the principal tensile stress criterion, such that 
\begin{equation}\label{eq:Dd}
    D_d = \zeta \left\langle \sum_{a=1}^3 \left( \frac{\langle \tilde{\sigma}_a \rangle}{\sigma_c} \right)^2 - 1 \right \rangle
\end{equation}

\begin{figure}[H]
    \centering
    \includegraphics[width=0.75\textwidth]{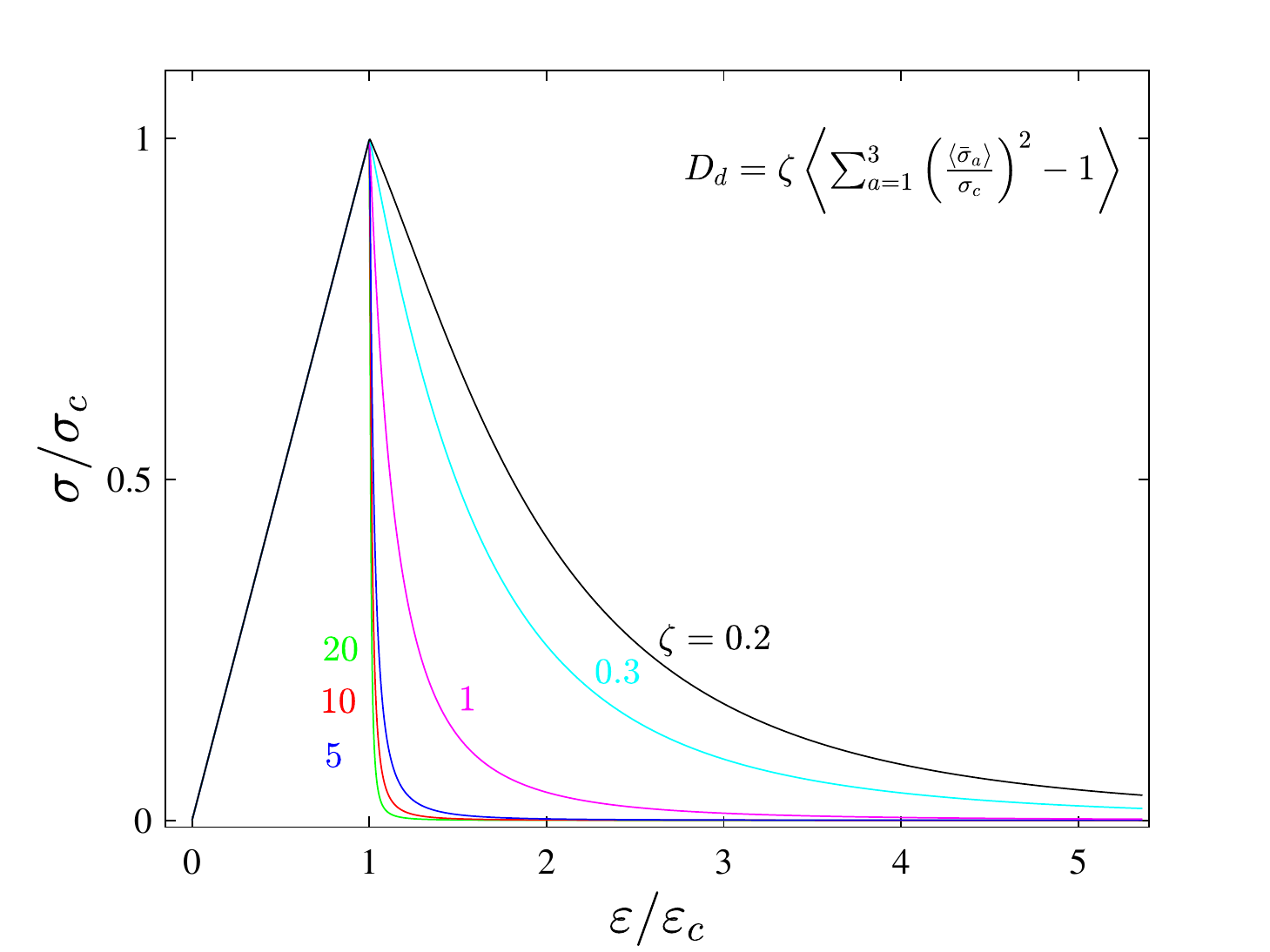} 
    \caption{Uniaxial stress-strain response, as predicted by the stress-based phase field model adopted, showcasing the role of the post peak parameter $\zeta$ on the material's post failure behaviour.}
    \label{fig:PPSP_Stress_Strain}
\end{figure}

Such a crack driving force state function is adequate for fractures resulting from the decohesion of surfaces perpendicular to the maximum principal stress and provides a quadratically increasing stress threshold for stress levels above a failure surface in the principal stress space, as determined by the material strength $\sigma_c$. Also, Eq. (\ref{eq:Dd}) provides a criterion independent of the phase field length scale $\ell$, which minimises the sensitivity of the results to this parameter. Given that the finite element mesh has to be sufficiently fine to resolve $\ell$, typically requiring an element size seven times smaller \cite{CMAME2018}, this facilitates tackling the large scales inherent to iceberg calving. For completeness, a non-dimensional parameter $\zeta$ has been introduced that, for $\zeta \neq 1 $ values, influences the slope of the stress-strain curve in the post-critical range. This is shown below by exploring the one-dimensional predictions of (\ref{eq:PhiForceBalance}) and (\ref{eq:Dd}). Hence, the evolution of the phase field in a one-dimensional setting ($\nabla \phi=0$) is given by,
\begin{equation}
    \phi = \frac{2 D_d}{1+2D_d} \, ,
\end{equation}

\noindent and accordingly the damaged (homogenised) uniaxial stress is found by making use of the following relationship,
\begin{equation}
    \sigma = \left(1 - \phi \right)^2 \sigma_0 = \left( 1 - \frac{2 D_d}{1+2D_d} \right)^2 E \varepsilon 
\end{equation}


\noindent where $\varepsilon$ is the uniaxial strain. The responses obtained are shown in Fig. \ref{fig:PPSP_Stress_Strain}, for selected choices of the parameter $\zeta$. A linear response is predicted until the critical fracture stress is reached, with the post-critical regime being sensitive to the value of $\zeta$; higher values translate into a less dissipative damage process, with the response appearing to converge for $\zeta > 5$. For simplicity, we will assume $\zeta=1$ but will also consider its influence in a parametric study. 

\subsubsection{A porodamage description of meltwater-driven crevasse growth}

Meltwater plays a key role in crevasse propagation, introducing local tensile stresses that can become equal or larger than the lithostatic compressive stress. It is thus pivotal to incorporate the role of the water pressure $p_w$ in the damaged ($\phi=1$) and transition ($0<\phi<1$) regions, as meltwater can accumulate in damaged zones and in the localised pore structure that arises in the transition region due to the nucleation, growth and coalescence of microvoids and microcracks. To this end, we follow Terzaghi's concept of an effective stress \cite{Terzaghi1923} and Biot's theory of poroelasticity \cite{Biot1941}. Hence, the resulting stress tensor is defined as,
\begin{equation} \label{eqn:Poromechanics}
    \tilde{\bm{\sigma}} = (1-\phi)^2 \bm{\sigma}_0 - \left[1-(1-\phi)^2\right] p_w \alpha \boldsymbol{I} \, ,
\end{equation}

\noindent where $\alpha$ is Biot's coefficient. In this work, $\alpha=1$. The use of degradation functions in Eq. (\ref{eqn:Poromechanics}) constrains the water pressure to damaged regions and removes the load carrying capacity of ice in fractured domains. Here, the water pressure $p_w$ is a hydrostatic term that is depth dependent. For surface crevasses it is defined as, 
\begin{equation} \label{eqn: Surface Meltwater}
   p_{w} =  \rho_wg\left<h_s-\left(z-z_s\right)\right> \,,
\end{equation}

\noindent where $\rho_w$ is the density of freshwater, $h_s$ is the meltwater depth, $z$ is the vertical height and $z_s$ is the distance between the glacier base and the bottom of the crevasse (see Fig. \ref{fig:Phase_Field_Ice_Diagram}). The presence of the Macaulay brackets in Eq. (\ref{eqn: Surface Meltwater}) implies that the pressure is zero above the water surface. Also, it is important to note that $z_s$ is updated for every time increment, as defined by the minimum depth at which $\phi=1$. Consequently, the role of meltwater pressure extends beyond the initial damage zone and appropriately evolves with the propagating crevasse. On the other hand, for basal crevasses it is assumed that the crevasse is fully saturated with ocean-water at depths below the ocean-water level $h_w$. The water pressure within basal crevasses is then given by
\begin{equation}
    p_{w} = -\rho_s g \left< h_w -z \right>
\end{equation}

In this context, the material density is interpolated as a function of the damage state, and the freshwater ($\rho_w$) and glacial ice ($\rho_i$) densities, reading
\begin{equation}
    \rho = \left( 1 - \phi \right)^2 \rho_i + \left[ 1 - (1- \phi)^2 \right] \rho_w \, .
\end{equation}

\subsection{Finite Element implementation}
\label{Sec:FEM}

Finally, we proceed to formulate the particularised coupled balance equations and briefly describe the finite element implementation. Considering the constitutive choices described in Section \ref{Sec:ConstitutiveTheory}, the local force balances are given by,
\begin{equation}\label{eq:DispStrongForm}
    \nabla \cdot \left\{ (1-\phi)^2 \bm{C}_0 \left( \bm{\varepsilon} - \bm{\varepsilon}^v \right) - \left[1-(1-\phi)^2\right] p_w \boldsymbol{I} \right\} + \mathbf{b}= \rho \Ddot{\mathbf{u}} \,\,\,\,\,\,\,\, \text{in} \,\,\,\,\,\, \Omega
\end{equation}
\begin{equation}\label{eq:PhiStrongForm}
    \phi - \ell^2 \nabla^2 \phi = 2 \left( 1 - \phi \right) \max_{\tau \in[0,t]}  \zeta \left\langle \sum_{a=1}^3 \left( \frac{\langle \tilde{\sigma}_a \rangle}{\sigma_c} \right)^2 - 1 \right \rangle \,\,\,\,\,\,\,\, \text{in} \,\,\,\,\,\, \Omega
\end{equation}

\noindent with the natural boundary conditions
\begin{equation}
    \tilde{\bm{\sigma}} \cdot \mathbf{n} = \mathbf{T} \,\,\,\,\,\,\,\, \text{in} \,\,\,\,\,\, \partial \Omega
\end{equation}
\begin{equation}
    \nabla \phi \cdot \mathbf{n} = 0 \,\,\,\,\,\,\,\, \text{in} \,\,\,\,\,\, \partial \Omega
\end{equation}

\noindent Here, $\bm{C}_0$ is the elastic stiffness tensor and the ansatz in the right hand side of Eq. (\ref{eq:PhiStrongForm}) is introduced to ensure damage irreversibility. The discretised system resulting from the weak form of (\ref{eq:DispStrongForm})-(\ref{eq:PhiStrongForm}) is solved using a so-called multi-pass (alternate minimization) staggered scheme \cite{Miehe2010a}. An implicit BDF time-stepping scheme is employed to solve, in a Backward Euler fashion, each set of equations. The commercial finite element package \texttt{COMSOL} is used. 

\section{Results}
\label{Sec:Results}

In this section, we present a series of 2D and 3D numerical examples, aimed at capturing the propagation of surface and basal crevasses within grounded glaciers and floating ice shelves. For 2D examples, we consider an idealised rectangular glacier of length $L=500$ m and height $H=125$ m, under the assumption of plane strain conditions. For simplicity, we neglect lateral shear and restrict the domain to a flow line near the terminus with $x$ and $z$ representing the along-flow and vertical coordinates. Gravitational load due to self-weight is applied as a uniform body force in the $z$-direction with a magnitude of $-\rho_i g$. We also consider the surface meltwater pressure $p_w$ within a crevasse using the poro-damage approach presented in Eq. (\ref{eqn: Surface Meltwater}). A Neumann-type traction is applied normal to the ice-ocean interface at the terminus, with the hydrostatic ocean-water pressure varying linearly with depth and a magnitude of $-\rho_s g \left< h_w -z \right>$. Boundary conditions that are specific to the grounded glacier and floating ice shelf cases are discussed in sections \ref{Sec:Grounded} and \ref{Sec:Floating}, respectively. Our simulations deal with glacial ice, whose material properties are given in Table \ref{tab: Properties2}, along with the densities of seawater and meltwater.

\begin{table} [H] 
    \centering
    \begin{tabular}{c c}
    \hline
       Material parameter &  Magnitude \\
       \hline
        Young's modulus, $E$ [MPa] & 9500 $^{\text{\cite{Karr1989AIce}}}$ \\
        Poisson's ratio, $\nu$ [-] & 0.35 $^{\text{\cite{Karr1989AIce}}}$ \\
        Density of glacial ice, $\rho_i$ [$\text{kg}/\text{m}^3$] & 917 $^{\text{\cite{Jimenez2018OnMechanics}}}$ \\
        Density of meltwater, $\rho_w$ [$\text{kg}/\text{m}^3$] & 1000 $^{\text{\cite{Jimenez2018OnMechanics}}}$ \\
        Density of seawater, $\rho_s$ [$\text{kg}/\text{m}^3$] & 1020 $^{\text{\cite{Jimenez2018OnMechanics}}}$ \\
        Fracture toughness, $K_{c}$ [$\text{MPa}\sqrt{\text{m}}$] & 0.10 $^{\text{\cite{Fischer1995}}}$ \\
        Critical fracture stress, $\sigma_c$ [MPa] & 0.1185 $^{\text{\cite{Krug2014CombiningCalving}}}$ \\
        Creep exponent, $n$ [-] & 3 $^{\text{\cite{Duddu2020}}}$ \\
        Creep coefficient $A$ [$\text{MPa}^{-\text{n}}\text{s}^{-1}$]  & 7.156 $\times 10^{-7}$ $^{\text{\cite{VanDerVeen2013FundamentalsDynamics}}}$ \\
         \hline
            \end{tabular}
    \caption{Material properties assumed in this work (unless otherwise stated). The values are chosen to characterise the behaviour of glacial ice, with the subscript number denoting the relevant reference.}
    \label{tab: Properties2}
\end{table}

The strength $\sigma_{c}$ magnitude is chosen to be an intermediate magnitude within the experimentally reported values of the critical fracture stress in glacial ice, which are in the range 0.08-0.14 MPa \cite{Fischer1995, Rist1999, Rist1996ExperimentalResults}. An estimate of the phase field length scale, which plays a negligible role in this model, can be obtained through the Hillerborg \textit{et al.} \cite{Hillerborg2008AnalysisElements} relation, which for plane strain reads: $\ell=(1-\nu^2)K_c^2/\sigma_c^2$. Considering the toughness of glacial ice ($K_c=0.1$ MPa$\sqrt{\text{m}}$), this gives a magnitude of $\ell=0.625$ m, which is the value adopted here (unless otherwise stated). To attain mesh-independent results, the characteristic element size along the crevasse propagation region is always chosen to be at least 5 times smaller than the phase field length scale $\ell$. 

\subsection{Propagation of a single crevasse on a grounded glacier}
\label{Sec:Grounded}

\begin{figure} [H]
\begin{subfigure}{\textwidth}
  \centering
  \includegraphics[width = 0.95\textwidth]{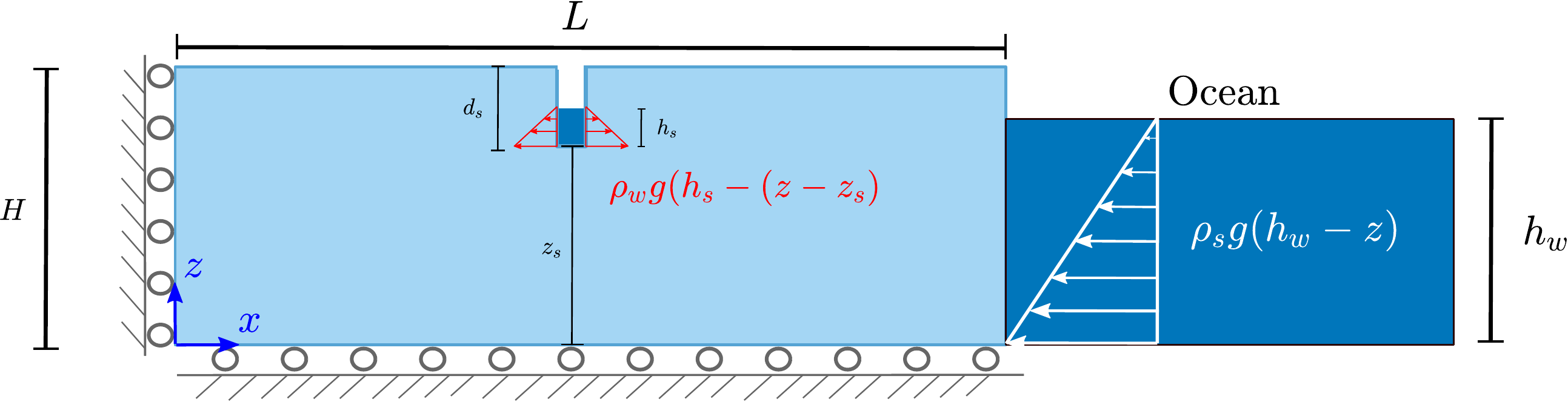} 
  \caption{}
  \label{fig:Grounded BCs}
\end{subfigure}
\begin{subfigure}{\textwidth}
  \centering
  \includegraphics[width = 0.75\textwidth]{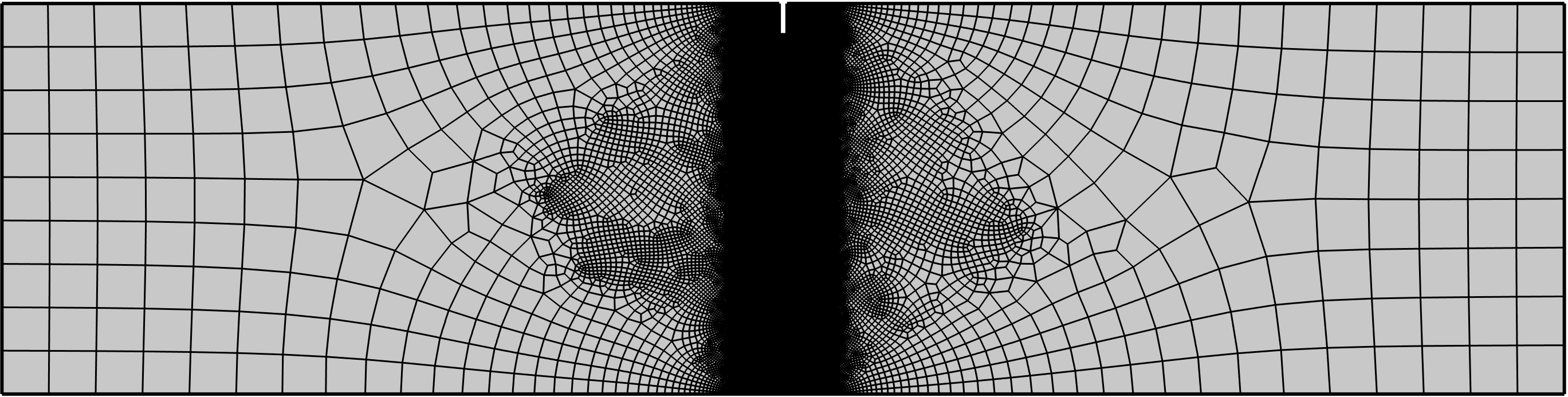} 
  \caption{}
  \label{fig:Mesh}
\end{subfigure}
\caption{Crevasse growth in a grounded glacier: (a) diagram showing the boundary conditions of a grounded glacier containing a single surface crevasse, and (b) finite element mesh employed, with the mesh refined along the expected crevasse propagation path.}
\label{fig:Mesh + BCs}
\end{figure}

We begin our numerical experiments by gaining insight into the behaviour of crevasses in grounded glaciers. Mimicking the conditions relevant to grounded glaciers, a free slip condition is applied to the bottom surface, restraining the displacement in the vertical direction. The normal component of the displacement field at the far left edge is restrained to prevent rigid body motion in the horizontal direction. The top surface, representing the atmosphere-ice interface, is defined as a free boundary. A visual representation of the geometry and boundary conditions for the grounded glacier can be found in Fig. \ref{fig:Grounded BCs}. In each of the following simulations, we refine the mesh beneath the initial notch, seen in Fig. \ref{fig:Mesh}. The entire domain is discretised using approximately 200,000 quadrilateral quadratic elements. 

\subsubsection{Stress state within a pristine grounded glacier}

\begin{figure} [H]
\makebox[\textwidth][c]{\centering
    \centering
    \includegraphics[width=1.1\textwidth]{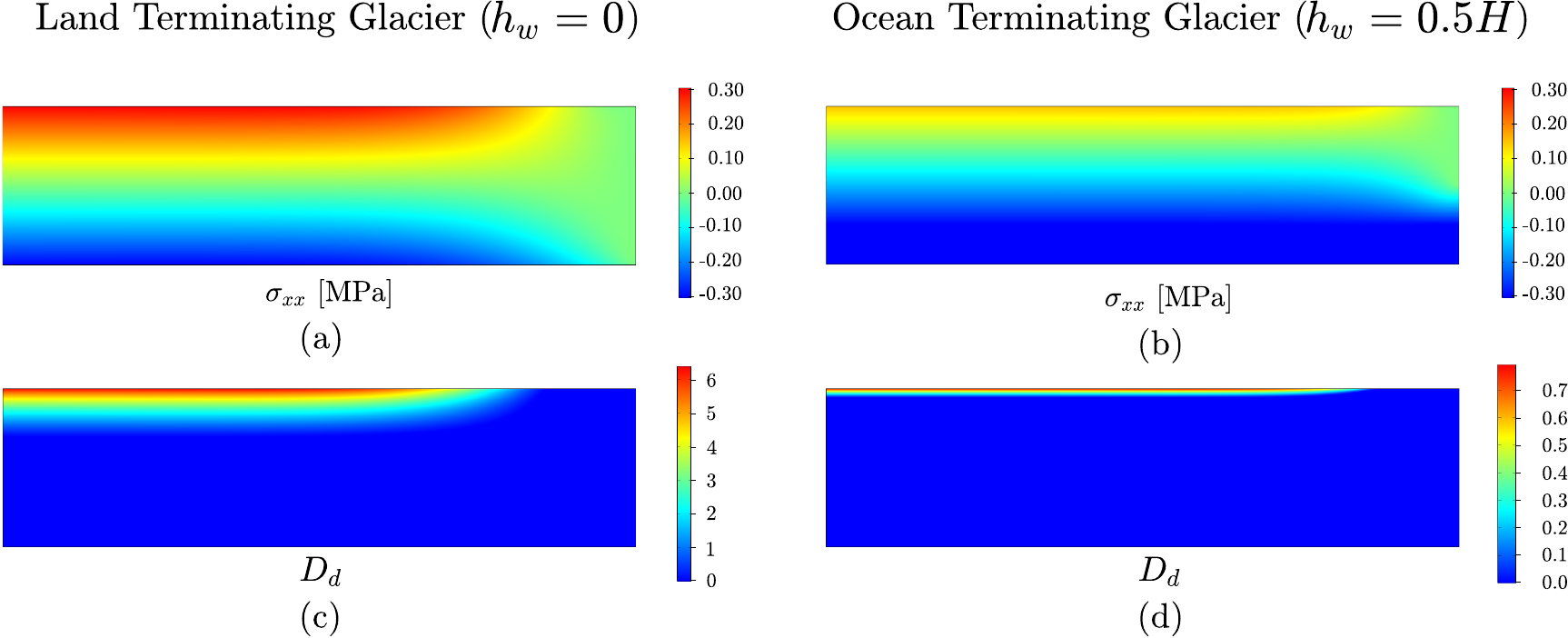}}
    \caption{Pristine grounded glacier. Contours of the longitudinal stress $\sigma_{xx}$, (a) and (b), and the crack diving force state function $D_d$, (c) and (d), for a land terminating glacier ($h_w=0$) and an ocean terminating glacier ($h_w = 0.5H$).}
    \label{fig:Stress States}
\end{figure}

Prior to introducing damage, we determine the stress states within pristine glaciers that are land terminating ($h_w = 0$) and ocean terminating ($h_w = 0.5H$). For simplicity, a linear elastic rheology is assumed. Important variables are the stresses in the longitudinal $x$-direction $\sigma_{xx}$, and the crack driving force $D_d$, given by Eq. (\ref{eq:Dd}). The results obtained are reported in Fig. \ref{fig:Stress States}, in terms of contours of $\sigma_{xx}$ and $D_d$. An edge effect on $\sigma_{xx}$ is observed at the far right terminus as a result of the traction free condition. However, away from the glacier terminus, the longitudinal stress field is invariant with the $x$-coordinate, owing to the idealised rectangular geometry. The maximum tensile stress occurs at the top surface and varies linearly with depth to a compressive region at the base, for both land and ocean terminating glaciers. For a land terminating glacier, the distribution of longitudinal stress is symmetric along the centre-line $z=H/2$, similar to the stress profile resulting from pure bending of a cantilevered beam \cite{Jimenez2018OnMechanics}. The effect of including the ocean-water pressure at the glacier terminus on the far field longitudinal stress can be observed by comparing Figs. \ref{fig:Stress States}a and \ref{fig:Stress States}b. Here, the ocean-water pressure provides a compressive stress that is constant with depth (in the far field region) and that decreases the extent of the tensile stress region near the top surface. If the ocean-water height is sufficiently large ($\approx$ 90\% of ice thickness), this can cause the glacier to become buoyant and form a floating ice shelf/tongue, resulting in an increased compressive stress regime. Vertical stress predictions (not shown) exhibit a behaviour that is also invariant with $x$-coordinate and that is compressive throughout the entire geometry, with the vertical stress being zero at the top surface and increasing linearly with depth. 

Consider now the crack driving force state function $D_d$ contours, Figs. \ref{fig:Stress States}c and \ref{fig:Stress States}d. Because only principal tensile stresses above the material strength contribute to damage, see Eq. (\ref{eq:Dd}), $D_d$ is only non-zero in the upper region. This agrees with the expected distribution for the damage driving force; non-zero in the tensile regions, with the maximum value located at the upper surface, and zero in regions of compressive stress. Since the vertical stresses are compressive throughout the entire profile, any crevasse propagation should be a mode I fracture, driven by the longitudinal stress $\sigma_{xx}$. Unlike strain energy based approaches \cite{Sun2021}, the present formulation appropriately captures a damage driving force that is only positive in tensile stress regions, consistent with linear elastic fracture mechanics (LEFM) predictions. 

\subsubsection{Crevasse propagation}
\label{Sec: Single Crevasse Growth}

We next consider a grounded glacier with an isolated surface crevasse, represented by an initial rectangular notch of height $d_s=2.5$ m and width $b=10$ m, which is located at mid-length of the top surface. This facilitates comparisons with LEFM. Following Ref. \cite{Sun2021}, we also consider a damage threshold $F^{\text{th}}$, below which $D_d=0$. As discussed in Ref. \cite{Sun2021} and shown below, this threshold has no influence on the final crevasse depth predicted but assists in localising damage. The magnitude of $F^{\text{th}}$ is chosen to be the maximum value of $D_d$ predicted in the pristine (unnotched) glacier simulation. In this way, one can ensure that damage only nucleates ahead of the crevasse, in agreement with the conditions relevant to the LEFM analysis (where crack nucleation does not occur). A similar effect can be achieved by increasing the value of the critical fracture stress $\sigma_c$. However, more research is needed before a quantitative link can be established between a material property and the damage threshold required to localise cracking ahead of the initial crevasse, as the latter appears to be dependent on the boundary value problem under consideration. We start the finite element analysis by initialising the stress state, in the absence of damage, and then conduct a subsequent time-dependent step to predict crevasse growth. The contributions from kinetic energy are found to play an important role in regularising the problem as, in the absence of inertia, equilibrium requires balancing an internal load carrying capacity that is being degraded by the damage with a prescribed gravity load. This suggests a deeper investigation into the role of inertia in ice-sheet fracture, which will be the objective of future work. In each simulation, the meltwater depth ratio $h_s/d_s$ is kept at a constant value (i.e. the meltwater depth increases proportionally with the crevasse depth). A parametric study is carried out for selected values of ocean-water level $h_w = (0, 0.5H, 0.9H )$ and meltwater depth ratios, to determine their influence on final crevasse depths. The results from the computational model are then compared with the stabilised crevasse depths predicted by LEFM using the `double edge cracks' weighting functions presented in \ref{Appendix B}. This study was performed for both linear elastic and non-linear viscous rheologies.

\subsubsection{Linear Elastic Rheology} \label{sub:Comparison with LEFM}

\begin{figure}[H]
\begin{subfigure}{.5\textwidth}
  \centering
    \includegraphics[height = 7cm]{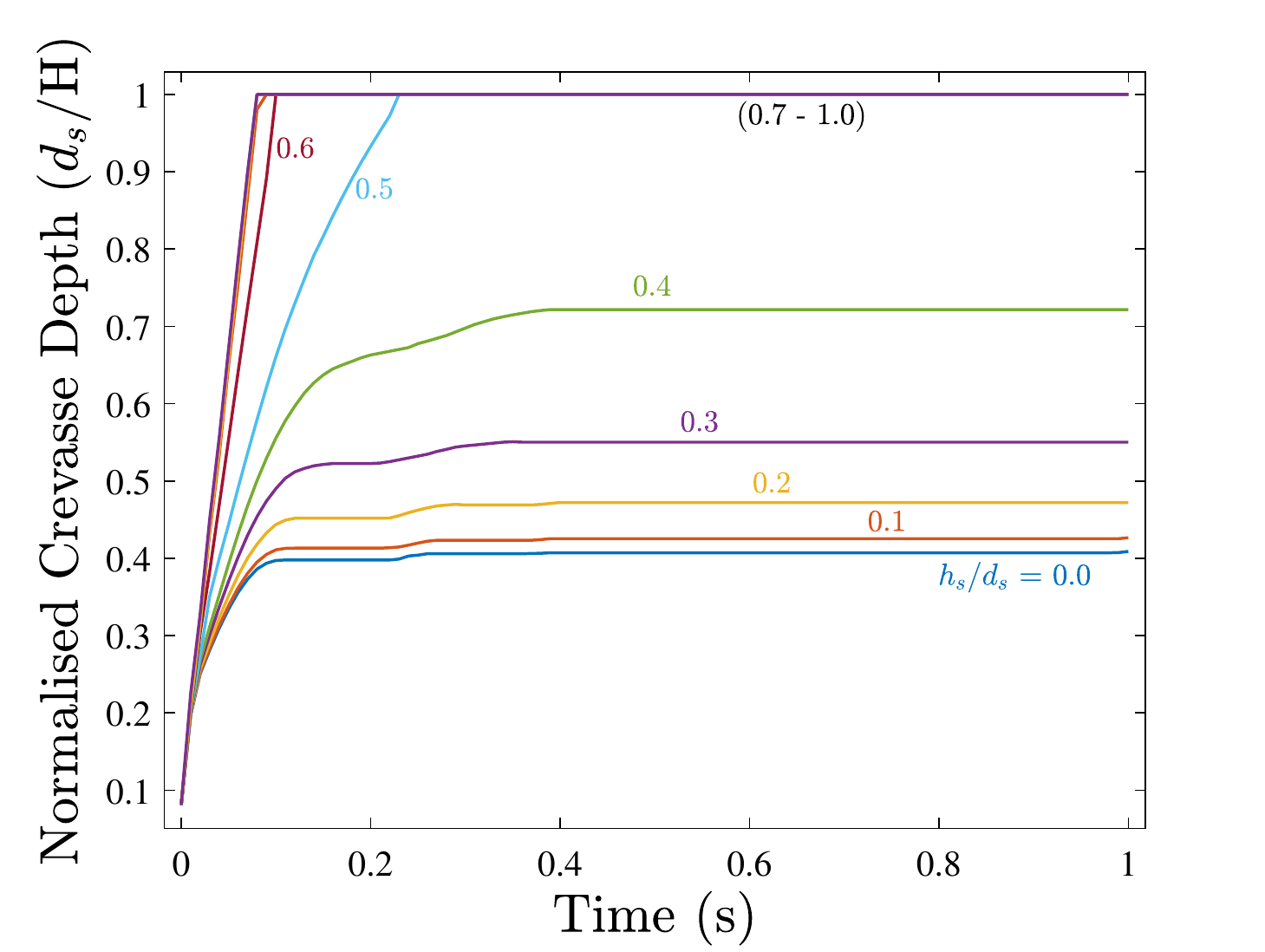}
    \caption{}
    \label{fig:Crevasse depth vs Time Linear Elastic}
    \end{subfigure}
\begin{subfigure}{.5\textwidth}
  \centering
    \includegraphics[height = 7cm]{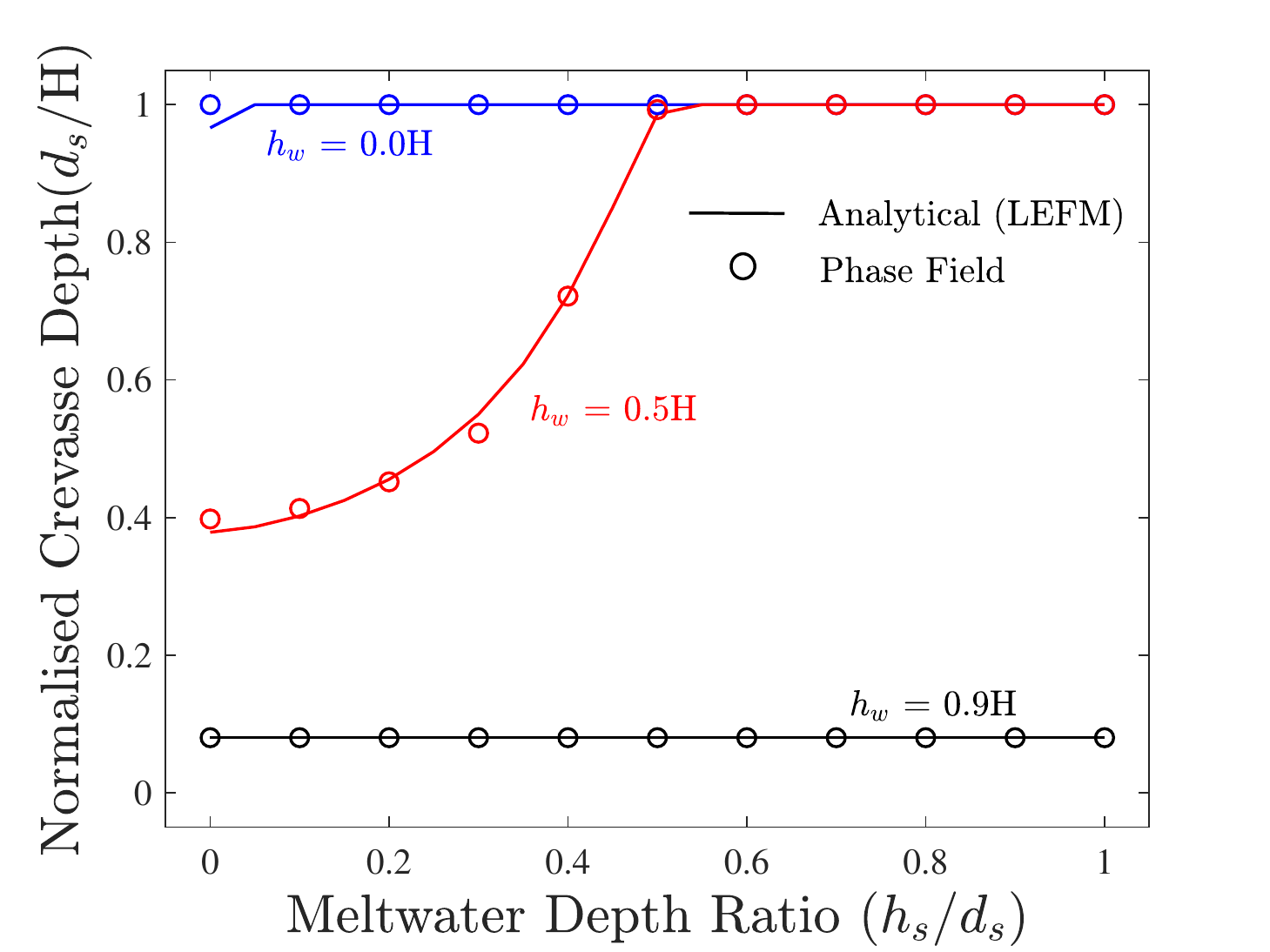}
    \caption{}
    \label{fig:PFMvsLEFM Elastic}
\end{subfigure}
    \caption{Crevasse growth in a grounded glacier. Normalised crevasse depth predictions for a single isolated crevasse in a linear elastic ice sheet: (a) phase field predictions of normalised crevasse depth versus time; and (b) phase field and analytical LEFM predictions of normalised crevasse depth versus meltwater depth ratio as a function of the ocean-water height.}
    \label{fig:Elastic Model}
\end{figure}

We first consider a linear elastic rheology for the grounded glacier, so as to validate model predictions with those obtained using analytical LEFM methods. The computational predictions of normalised crevasse depth versus time are shown in Fig. \ref{fig:Crevasse depth vs Time Linear Elastic} for an ocean-water height of $h_w = 0.5H$ and selected values of the meltwater depth. It can be seen that the crevasses propagate rapidly and stabilise at a constant depth. In agreement with expectations, larger meltwater depths lead to higher crevasse depths, with the crevasse propagating all the way to the base of the glacier for $h_s/d_s>0.5$. A plot of the normalised stable crevasse depths for both the analytical LEFM and phase field models is given in Fig. \ref{fig:PFMvsLEFM Elastic} as a function of the meltwater and ocean-water depth ratios. The stabilised crevasse depths estimated with the phase field model show a very good agreement with those predicted using LEFM for all values of meltwater depth ratio and ocean-water height. It can be seen that land terminating glaciers ($h_w = 0$) are susceptible to deeper fractures, even without the presence of meltwater, as there is no ocean-water compressive pressure at the terminus. The crevasse depth reduces significantly when ocean-water is present. For example, a dry crevasse is predicted to propagate to 37.8$\%$ of the glacier height for an ocean-water depth of $h_w = 0.5H$. Crevasse depth gradually increases with meltwater depth ratio for ratios less than 0.5, whereas the crevasse penetrates the full glacier thickness for meltwater depth ratios greater than 0.5. For the near floating glacier cases ($h_w = 0.9H$) the compressive stresses due to the ocean-water are significantly large enough to completely offset the tensile regions in the upper surface of glacier, and thus there is no meltwater depth ratio at which the crevasse can extend beyond the initial notch length. 

\begin{figure} [H]
\begin{subfigure}{\textwidth}
  \centering
  \includegraphics[width = 0.95\textwidth]{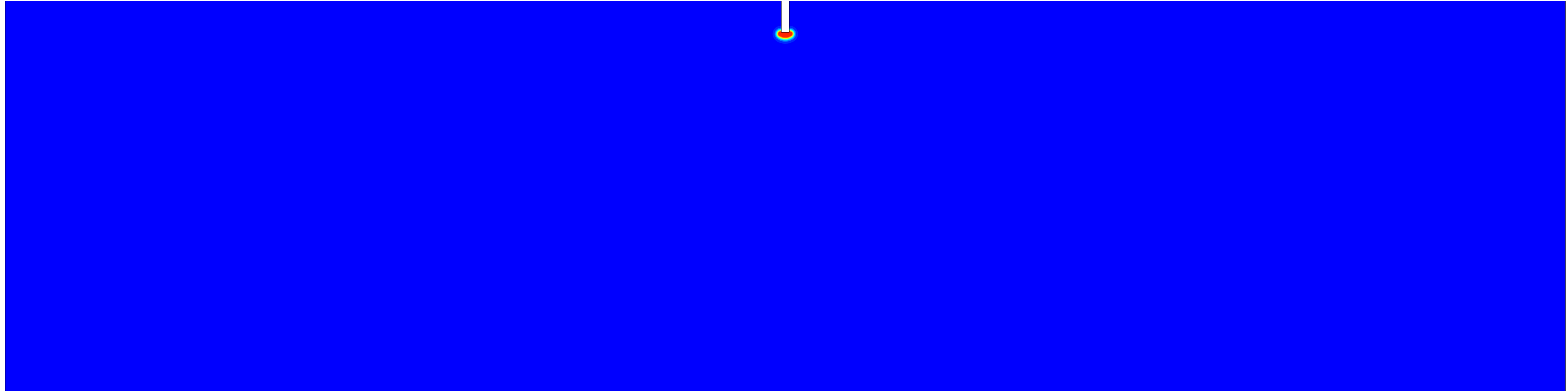}
  \caption{}
  \label{fig:Phase Field Meltwater 0.2 t = 0.00s}
\end{subfigure}
\begin{subfigure}{\textwidth}
  \centering
  \includegraphics[width = 0.95\textwidth]{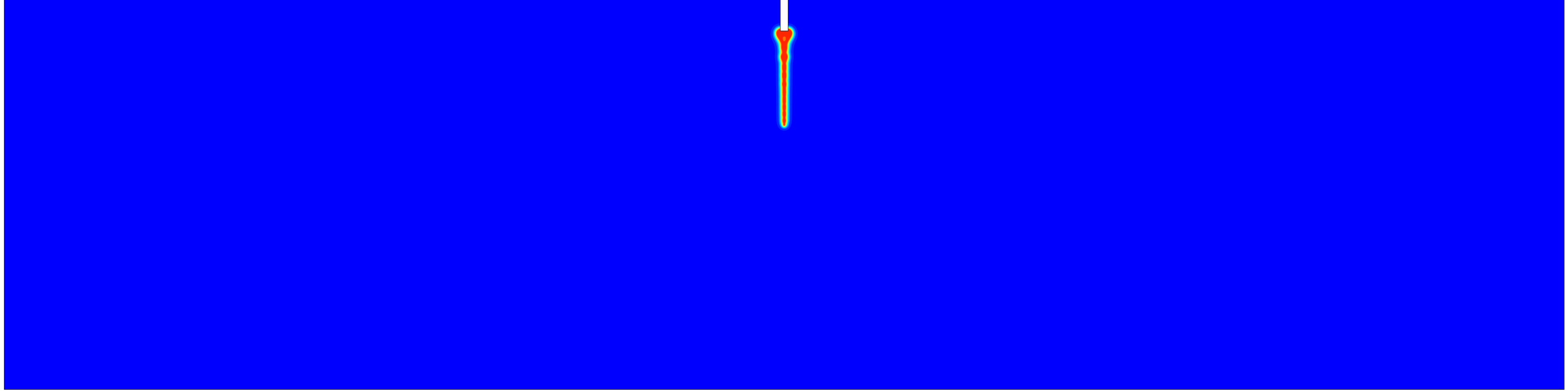}
  \caption{}
  \label{fig:Phase Field Meltwater 0.2 t = 0.05s}
\end{subfigure}
\begin{subfigure}{\textwidth}
  \centering
 \includegraphics[width = 0.95\textwidth]{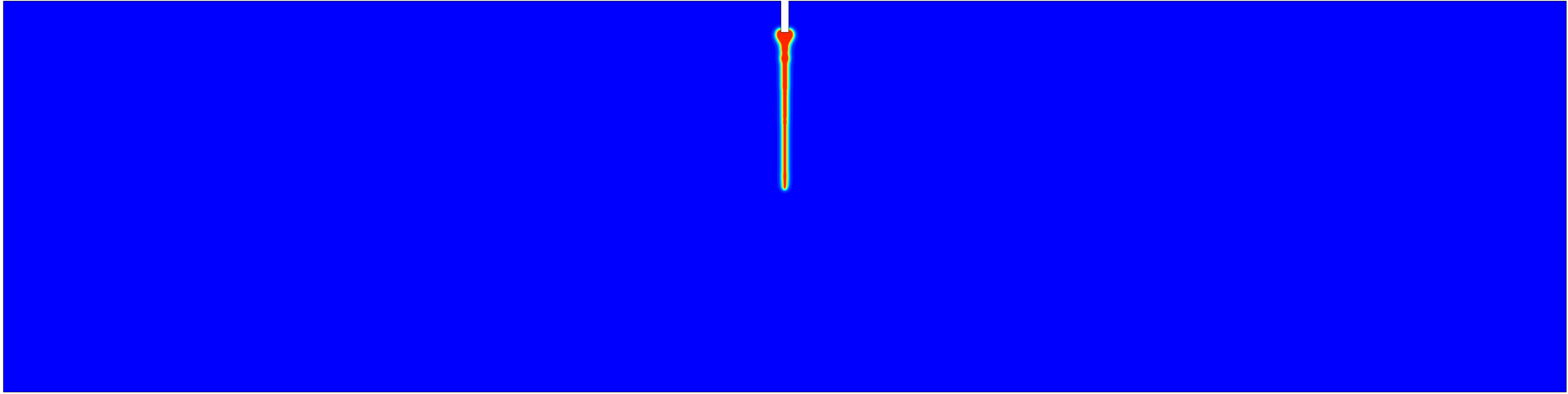} 
  \caption{}
  \label{fig:Phase Field Meltwater 0.2 t = 0.40s}
\end{subfigure}
\begin{subfigure}{\textwidth}
  \centering
 \includegraphics[width = 1.05\textwidth]{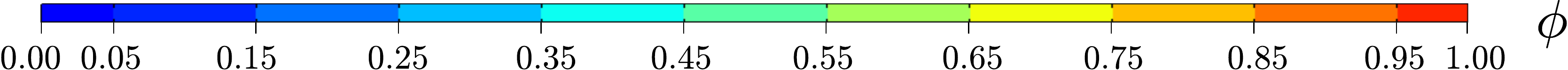} 
\end{subfigure}
\caption{Crevasse growth in a grounded glacier. Phase field damage evolution as a function of time: (a) t = 0.00 s, (b) t = 0.02 s, and (c) t = 0.40 s. The results correspond to the case of a meltwater depth ratio of $h_s/d_s = 0.2$ and an ocean-water height of $h_w = 0.5H$, assuming a linear elastic compressible rheology.}
\label{fig:Phase Field Meltwater 0.2}
\end{figure}

The process of crevasse growth is shown in Fig. \ref{fig:Phase Field Meltwater 0.2}, through plots of phase field $\phi$ contours at selected time intervals. The results correspond to the case of a meltwater depth ratio of $h_s/d_s = 0.2$ and an ocean-water height of $h_w = 0.5H$, but the qualitative behaviour is the same in all cases. A sharp mode I crack propagates directly below the initial crevasse until reaching the region where the compressive stresses are sufficiently large to arrest the crack.

\subsubsection{Parametric analysis} 
\label{Sub: Sensitivity Analysis}

\begin{figure} [H]
   \begin{subfigure} {0.5\textwidth}
     \centering
    \includegraphics[height = 6.25cm]{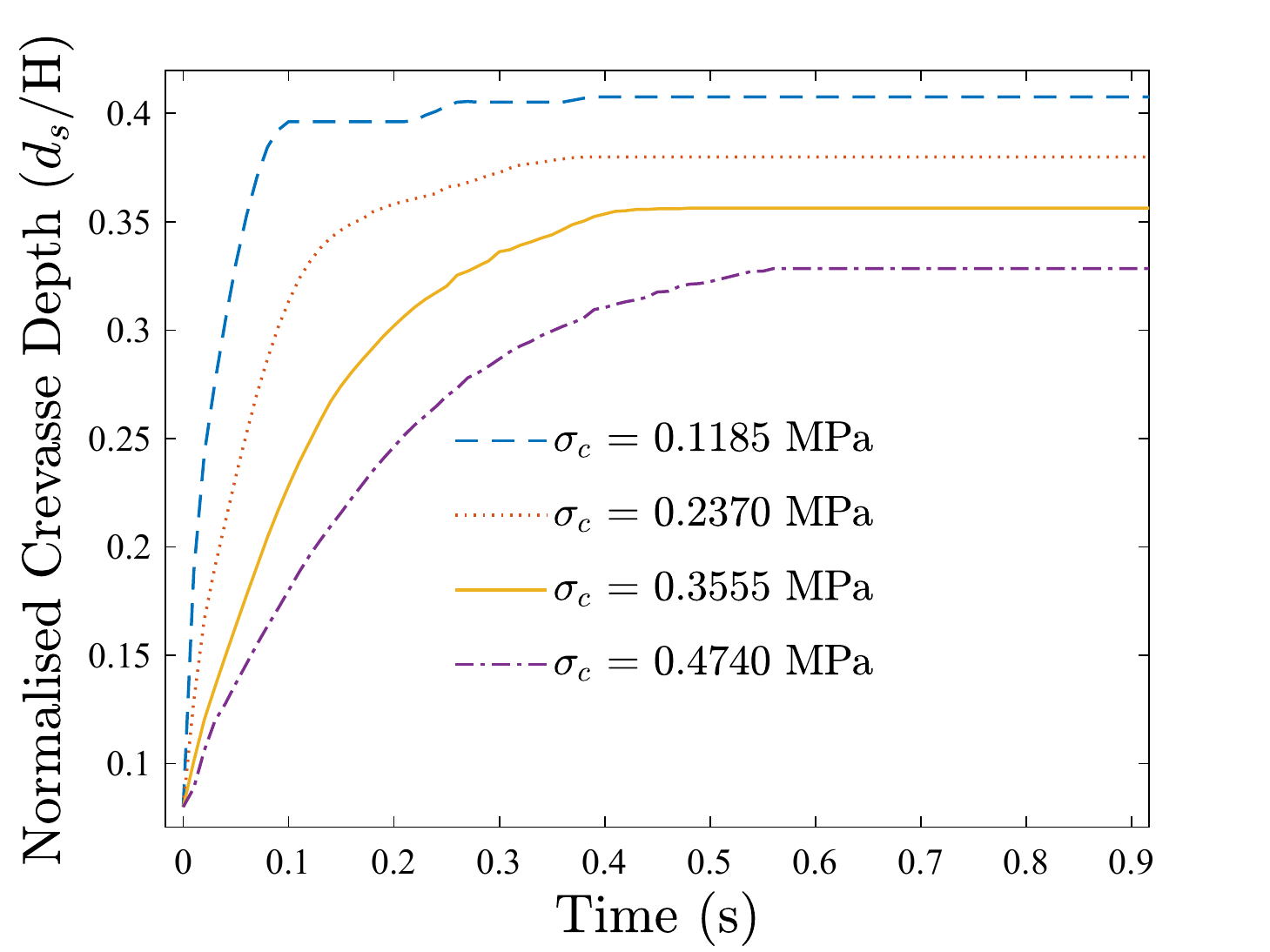}
    \caption{}
    \label{fig:Critical stress sensitivity}
   \end{subfigure}
   \hfill
   \begin{subfigure} {0.5\textwidth}
     \centering
    \includegraphics[height = 6.25cm]{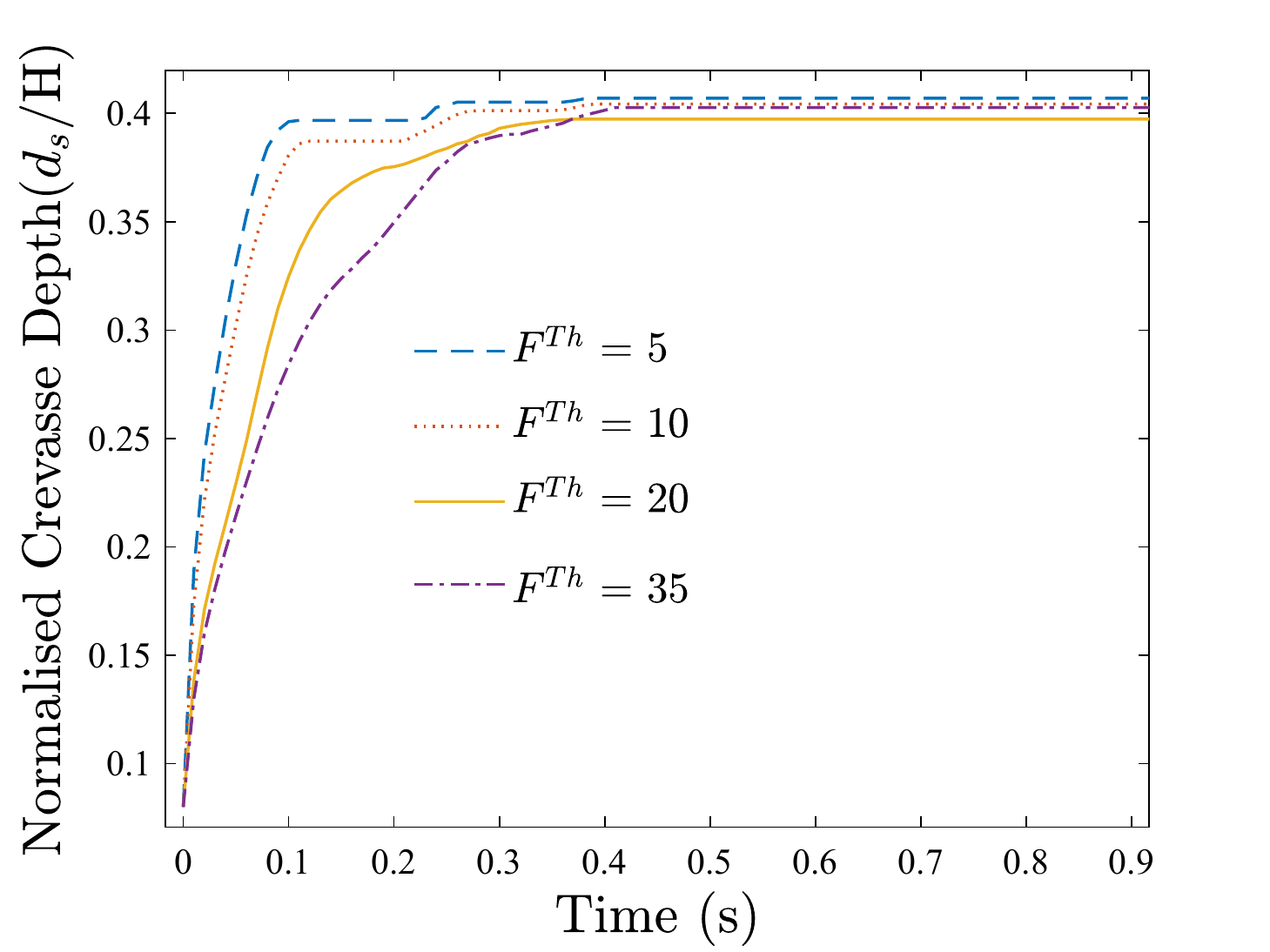}
    \caption{}
    \label{fig:Crack driving force threshold sensitvity}
   \end{subfigure}
     \hfill
   \begin{subfigure} {0.5\textwidth}
     \centering
    \includegraphics[height = 6.25cm]{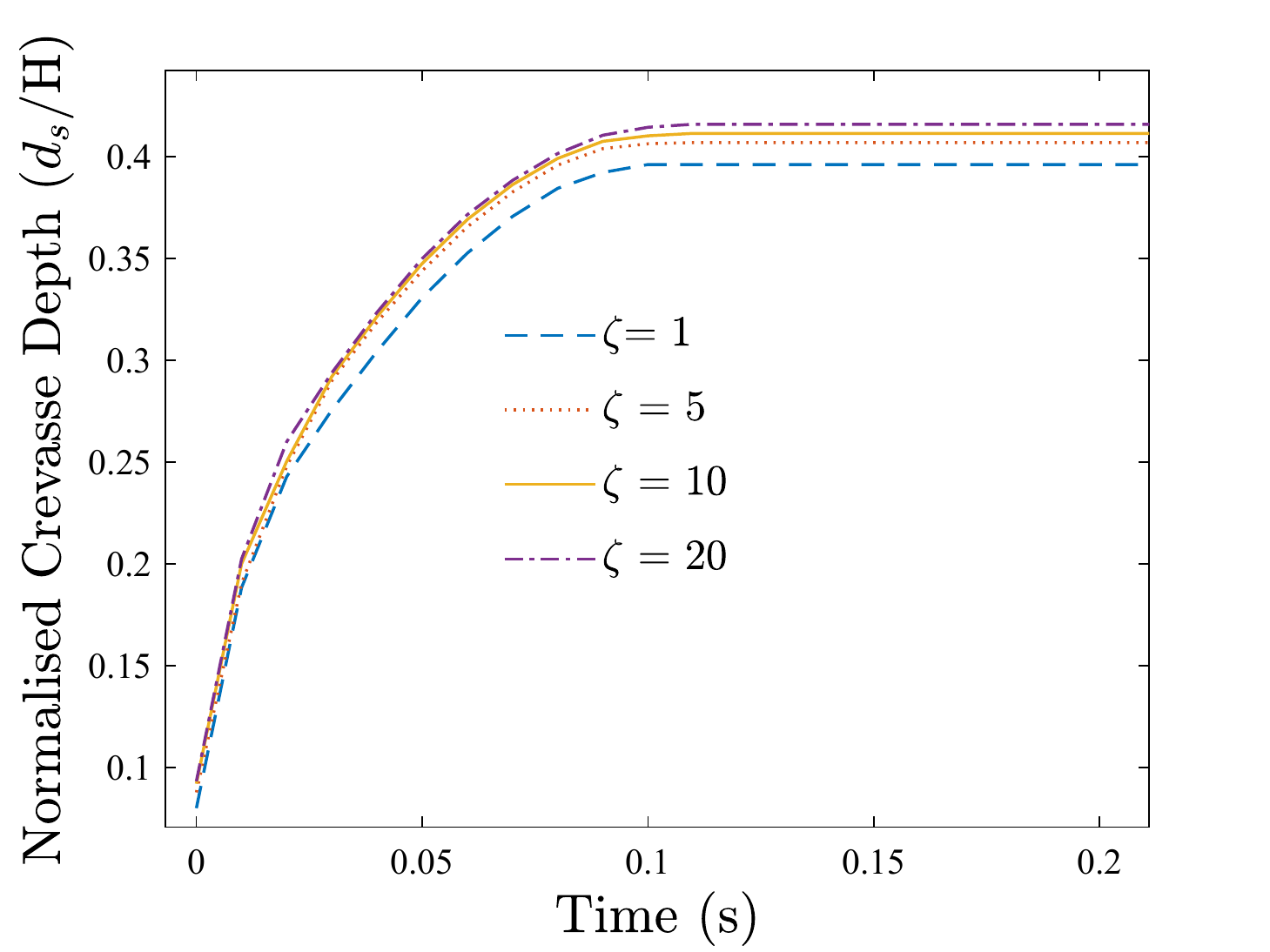}
    \caption{}
    \label{fig:Post peak slope parameter sensitivity}
   \end{subfigure}
   \begin{subfigure} {0.5\textwidth}
     \centering
    \includegraphics[height = 6.25cm]{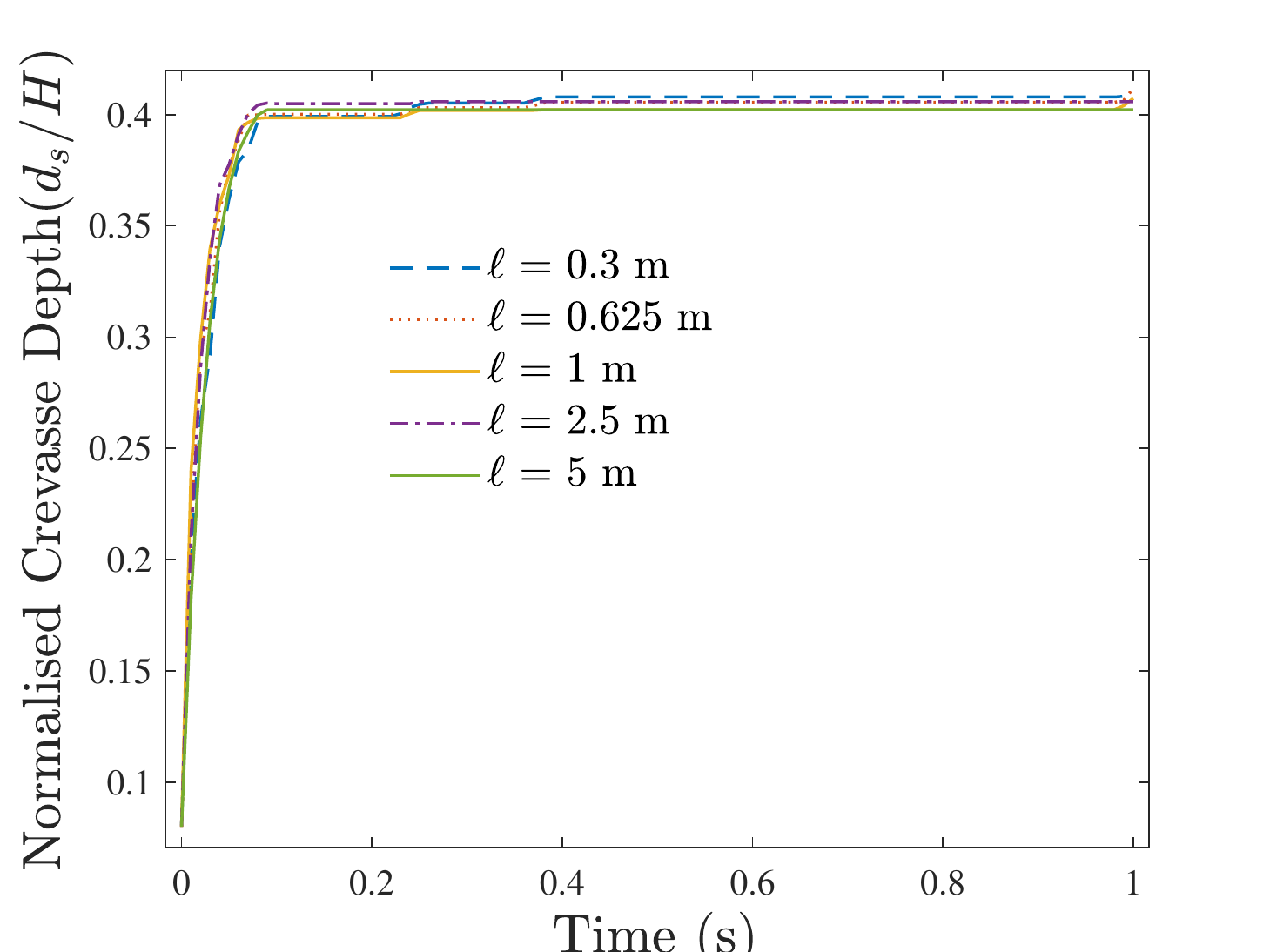}
    \caption{}
    \label{fig:Length scale parameter sensitivity}
   \end{subfigure}

\caption{Crevasse growth in a grounded glacier. Normalised surface crevasse depth versus time predictions for a dry isolated crevasse with an ocean-water ratio of $h_w = 0.5H$. Parametric study varying (a) critical fracture stress $\sigma_c$ , (b) crack driving force threshold $F^{\text{th}}$, (c) post peak slope parameter $\zeta$, and (d) phase field length scale $\ell$.}
\label{fig:Sensitivity}
   
\end{figure}

We shall now conduct sensitivity studies on relevant material, fracture and numerical parameters. The base model considered here is an isolated dry surface crevasse with an ocean-water level $h_w=0.5H$. We consider the individual effect on the stabilised crevasse depth of the mode I critical fracture stress or cohesive strength $\sigma_c$, the crack driving force threshold $F^{\text{th}}$, the post peak slope parameter $\zeta$, and the phase field length scale $\ell$, whilst keeping all other parameters constant. The results obtained are shown in Fig. \ref{fig:Sensitivity}. 

Consider first the sensitivity to the material strength $\sigma_c$, Fig. \ref{fig:Critical stress sensitivity}, which is varied within the range $0.1185 - 0.4740$ MPa. In agreement with expectations, the predicted crevasse depth decreases with increasing $\sigma_c$. The results obtained for different values of the crack driving force threshold can be found in Fig. \ref{fig:Crack driving force threshold sensitvity}. We find that there is little variation in predicted final crevasse depth when increasing the threshold to up to seven times, with a maximum percentage difference of $2.4\%$ between values of stabilised crevasse depth. The results obtained for various values of the post-peak parameter $\zeta$ are given in Fig. \ref{fig:Post peak slope parameter sensitivity}. A small influence is observed, with higher $\zeta$ values leading to larger crevasse depths, as they result in a higher $D_d$ magnitude for the same stress level. This is also consistent with the sharper drop in the uniaxial stress-strain curve with increasing $\zeta$ shown in Fig. \ref{fig:PPSP_Stress_Strain}. Finally, the sensitivity to the phase field length scale $\ell$ is explored in Fig. \ref{fig:Length scale parameter sensitivity}. The results confirm the rather negligible sensitivity of the phase field formulation employed to the magnitude of $\ell$.

\subsubsection{Non-linear viscous rheology}

\begin{figure}[H]
\begin{subfigure}{.5\textwidth}
  \centering
    \includegraphics[height = 7cm]{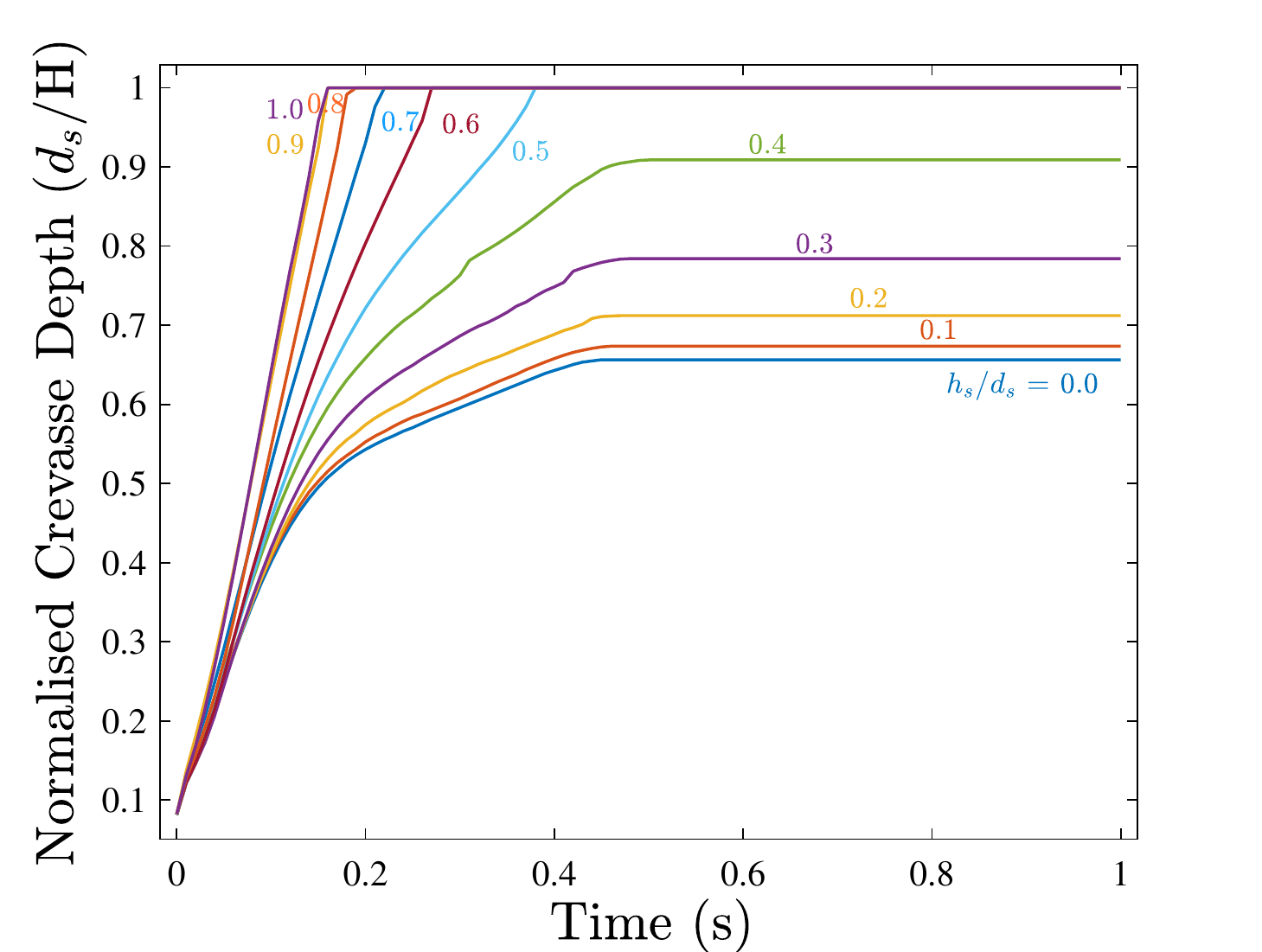}
    \caption{}
    \label{fig:Crevasse depth vs Time Linear Creep}
    \end{subfigure}
\begin{subfigure}{.5\textwidth}
  \centering
    \includegraphics[height = 7cm]{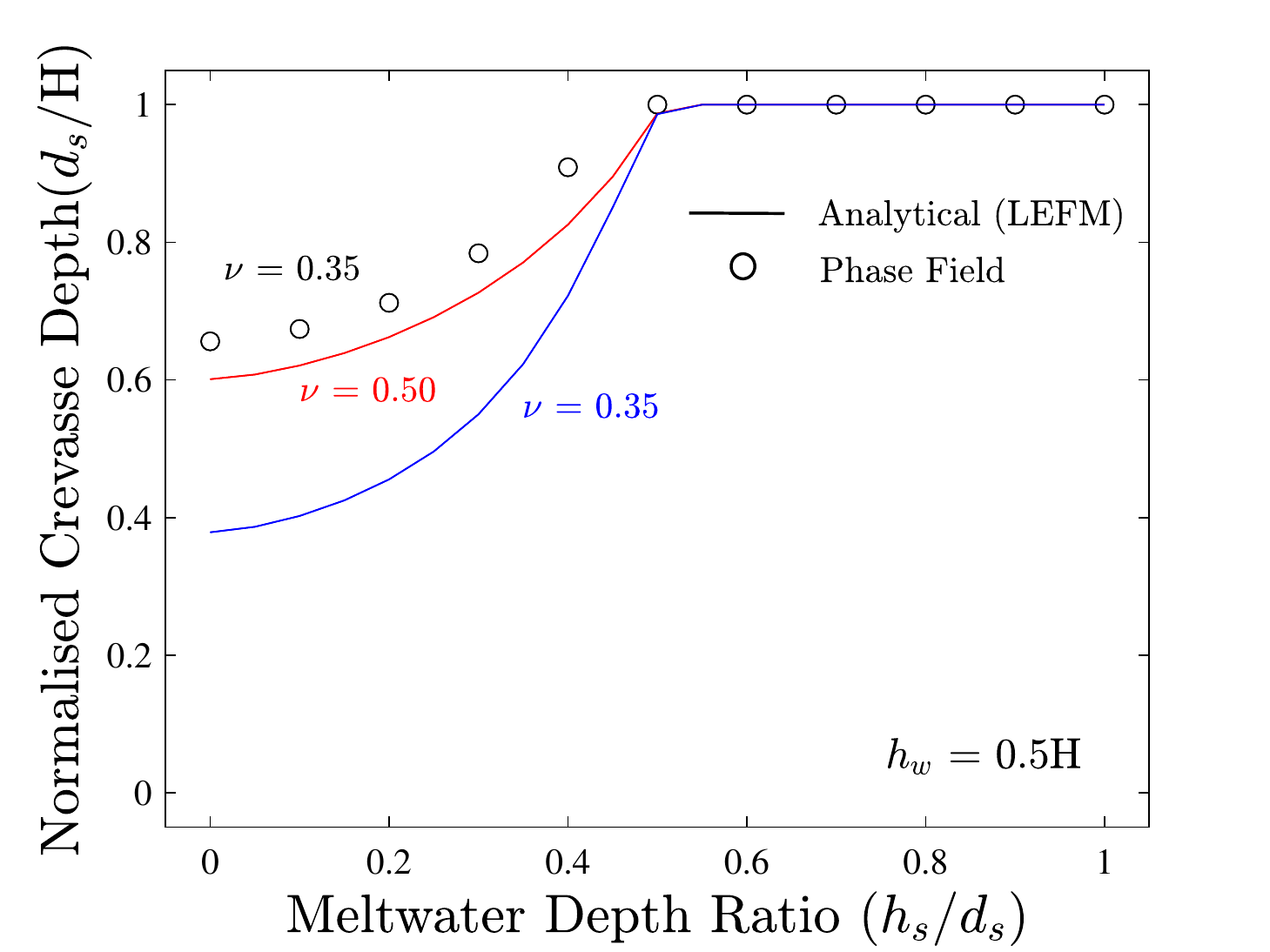}
    \caption{}
    \label{fig:PFMvsLEFM Creep}
\end{subfigure}
    \caption{Crevasse growth in a grounded glacier. Normalised crevasse depth predictions for a single isolated crevasse assuming a non-linear viscous rheology: (a) phase field predictions of normalised crevasse depth versus time; and (b) phase field and analytical LEFM predictions of normalised crevasse depth versus meltwater depth ratio as a function of the ocean-water height. The LEFM predictions are shown for both compressible ($\nu=0.35$) and incompressible ($\nu=0.5$) constitutive behaviour.}
    \label{fig:Creep Model}
\end{figure}

We next investigate the influence of the rheology upon the final crevasse depth by considering the non-linear viscous Glen's flow law (Section \ref{Sec:GlenFlowLaw}). Here, we run a time-dependent creep simulation without phase field damage to allow for a steady-state stress profile to develop within the glacier. The results of the creep simulation are then used to initialise the phase field model, so as to study the propagation of a crevasse based on an incompressible stress state. Results showing the normalised crevasse depth versus time for the non-linear viscous rheology are found in Fig. \ref{fig:Crevasse depth vs Time Linear Creep} with increasing values of meltwater depth ratio $h_s/d_s$ and for an ocean-water height of $h_w = 0.5H$. A comparison between the stabilised crevasse depths from the phase field model and LEFM can be found in Fig. \ref{fig:PFMvsLEFM Creep}. The influence of meltwater within the crevasse is qualitatively similar to the linear elastic case, with stabilised crevasse depths becoming progressively larger with increased meltwater. Full fracture is predicted at a meltwater depth ratio $h_s/d_s = 0.5$ or larger.

Consider now Fig. \ref{fig:PFMvsLEFM Creep}; two key observations emerge. First, neglecting the non-linear viscous rheology of ice implies underpredicting the extent of crevasse propagation. A dry glacier crevasse extends to $65.6\%$ of the glacier height when incorporating creep deformation, compared to only $37.8\%$ when considering a linear elastic compressive rheology. 
Second, the normalised crevasse depths from the phase field model (using $\nu=0.35$) are comparable to those from the LEFM model assuming incompressible behaviour ($\nu=0.5$). Despite the compressible elastic deformation, the longitudinal stress profile is dictated by the incompressible viscous deformation according to the Glen's law. Thus, we find that first-order estimates obtained from analytical LEFM approaches should consider a Poisson's ratio of $\nu=0.5$ to avoid underpredicting the impact of meltwater on ice-sheet stability. Our findings are consistent with the calculations by Plate \textit{et al.} \cite{Plate2012}, where Poisson's ratio was found to have a notable influence on the fracture driving force for elastic ice sheets.

\subsection{Propagation of multiple surface crevasses in a grounded marine-terminating glacier} 
\label{sub:Multiple Surface Crevasses}

\begin{figure} [H]
    \centering
    \includegraphics[width = \textwidth]{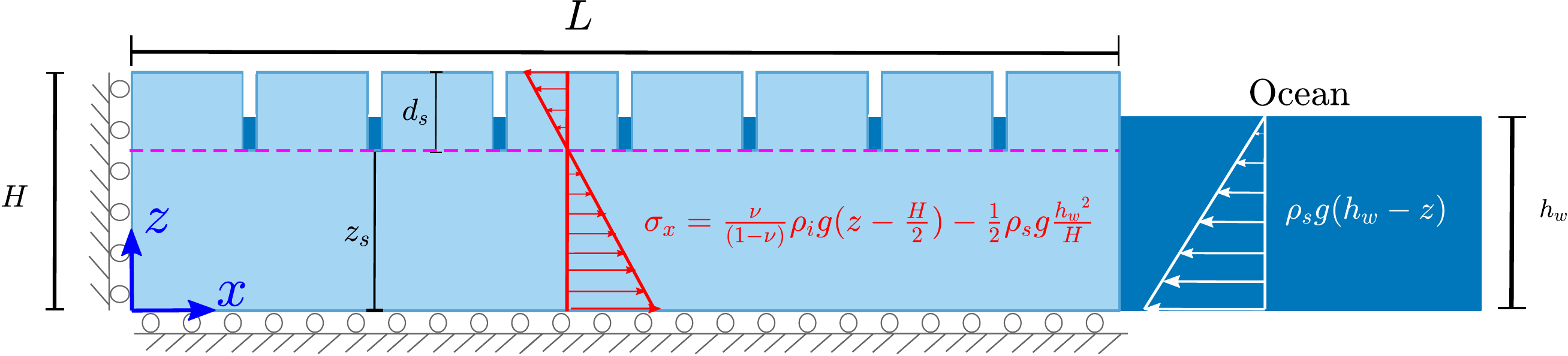}
    \caption{Multiple crevasse growth in a grounded marine-terminating glacier. Diagram showing the boundary conditions of a grounded glacier with a field of densely spaced crevasses (spaced 50 m apart from each other).}
    \label{fig:Multiple Surface Crevasse Diagram}
\end{figure}

We next determine the penetration depths for a uniform field of densely spaced surface crevasses. The same glacier geometry from the previous example is used, but we consider seven surface crevasses, each spaced at 50 m apart and located sufficiently far away from the glacier terminus, so that the edge effects do not influence crevasse growth (see Fig. \ref{fig:Multiple Surface Crevasse Diagram}). Here, we aim to study the effect of neighbouring crevasses, which are expected to provide crack shielding that reduces the final crevasse depth, and to compare the phase field model results with those predicted by the Nye zero stress model \cite{Nye1955CommentsCrevasses}. The results from the Nye zero stress model are found by computing the depth at which the far field longitudinal stress becomes zero, represented by the dashed purple line in Fig. \ref{fig:Multiple Surface Crevasse Diagram}. The model uses approximately 1.6 million linear triangular elements, with the mesh being refined ahead of each crevasse. 

\begin{figure} [H]
\begin{subfigure}{\textwidth}
  \centering
  \includegraphics[width = 0.95\textwidth]{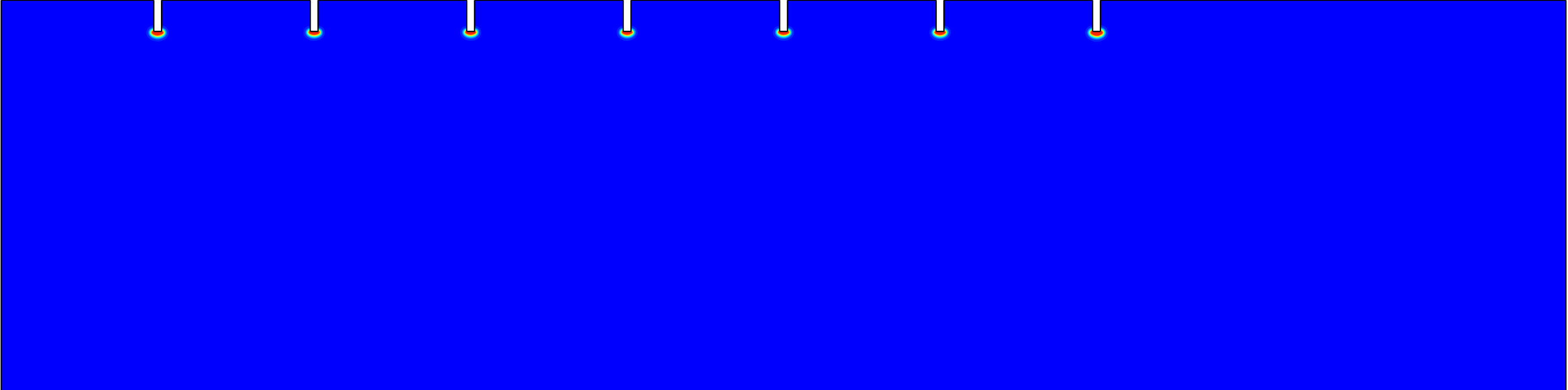}  
  \caption{}
  \label{fig:Multi Crevasse 0.00}
\end{subfigure}
\begin{subfigure}{\textwidth}
  \centering
  \includegraphics[width = 0.95\textwidth]{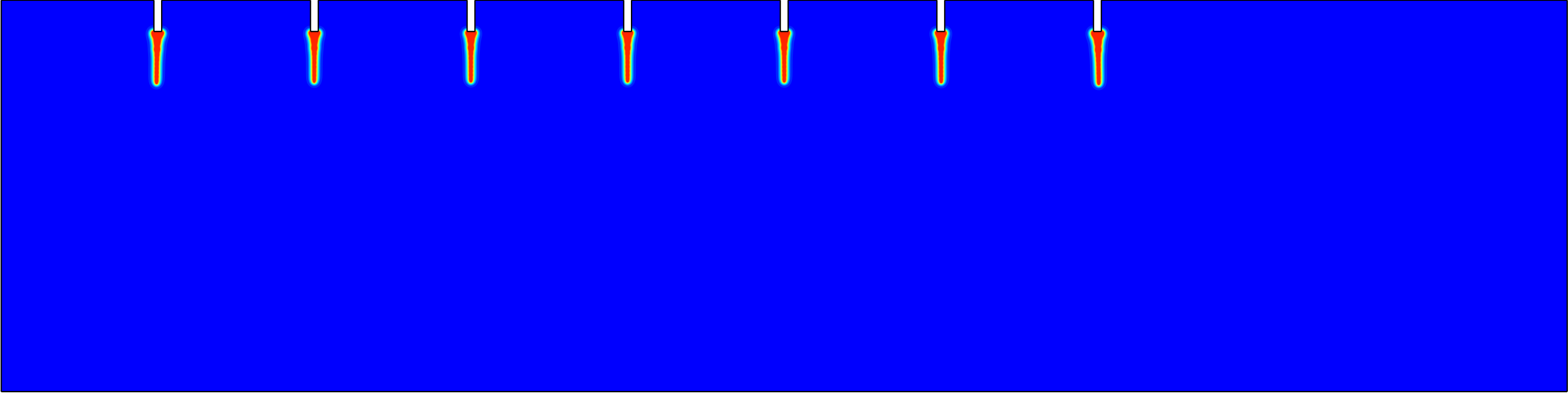} 
  \caption{}
  \label{fig:Multi Crevasse 0.01}
\end{subfigure}
\begin{subfigure}{\textwidth}
  \centering
  \includegraphics[width = 0.95\textwidth]{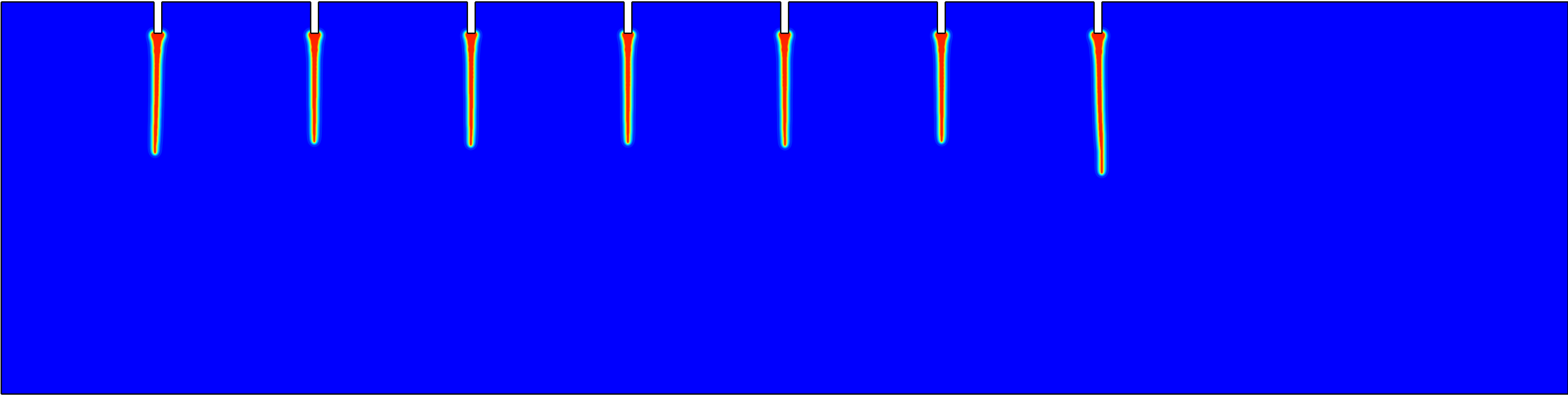} 
  \caption{}
  \label{fig:Multi Crevasse 0.40}
\end{subfigure}
\begin{subfigure}{\textwidth}
  \centering
  \includegraphics[width = 1.05\textwidth]{Phase_Field_Legend-eps-converted-to.pdf}
\end{subfigure}
\caption{Multiple crevasse growth in a grounded glacier. Phase field damage evolution as a function of time: (a) t = 0.00 s, (b) t = 0.01 s, and (c) t = 0.40 s. The results correspond to the case of a meltwater depth ratio of $h_s/d_s = 0.1$ and an ocean-water height of $h_w = 0.5H$, assuming a linear elastic compressible rheology.}
\label{fig:Multi Crevasse}
\end{figure}

Plots of the phase field damage variable can be found in Fig. \ref{fig:Multi Crevasse}, for an ocean-water height of $h_w = 0.5H$ and a meltwater depth ratio of $h_s/d_s = 0.1$. Qualitatively, the behaviour resembles that of the single crevasse model - crevasses propagate rapidly and subsequently arrest upon reaching the compressive region at the bottom. Each crevasse stabilises to a similar depth, although the outer crevasses penetrate slightly deeper because they experience shielding only from one side.

\begin{figure} [H]
    \centering
    \includegraphics[width = 0.75\textwidth]{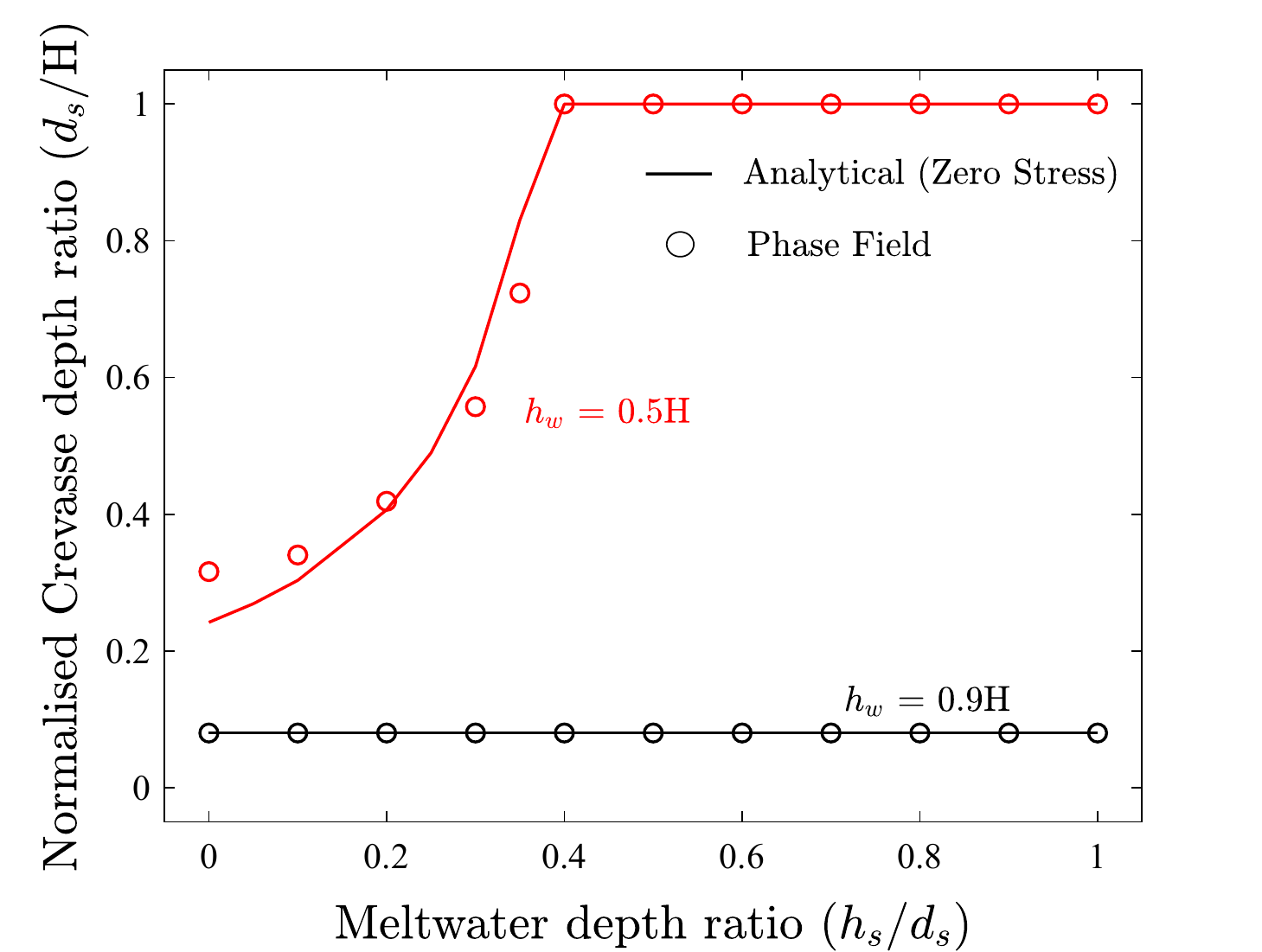}
    \caption{Multiple crevasse growth in a grounded glacier. Normalised crevasse depth versus meltwater depth ratio predictions as a function of the ocean-water height. Comparisons between the present phase field model and analytical predictions from Nye's zero-stress model \cite{Nye1955CommentsCrevasses}, for a linear elastic ice sheet.}
    \label{fig:ZSMvsPFMMulti}
\end{figure}

To shed light on the effect of crack shielding, we take measurements from the fourth crevasse at mid-length and compare with the predictions from the zero stress model; the results are shown in Fig. \ref{fig:ZSMvsPFMMulti}. The agreement is overall very good; as also observed in the LEFM comparisons, the model provides a good agreement with analytical predictions when particularised to the conditions where these analytical estimates are relevant. For the specific case of ocean-water height of $h_w= 0.5H$, the phase field model predicts a slightly deeper crevasse penetration compared to the zero stress model for smaller values of meltwater depth ratio. For the near floating condition ($h_w = 0.9H$), the ocean-water height is sufficiently large to completely offset the tensile region in the upper surface of the glacier. Thus the longitudinal stress profile is compressive throughout the entire height of the glacier (except near the terminus) and no amount of meltwater in the crevasse can extend it beyond its initial geometry.

\subsection{Propagation of surface and basal crevasses on a floating ice shelf}
\label{Sec:Floating}

Ice shelves form along coastal regions of Antarctica as a result of ongoing glacial flow and associated thinning to the point at which grounded ice becomes afloat (i.e. the grounding line). Here there are two possibilities: (1) the mass loss terms at the grounding line (calving and melting) are greater than or equal to the flux of ice across the grounding line, and so the ice sheet will terminate here; and (2) the flux of ice exceeds mass loss terms, and ice flows across the grounding line to form a floating slab of ice. In this section, we assume plane strain conditions and consider a floating ice shelf of length $L = 5000$ m and height $H = 125$ m. To enforce the floating boundary condition at the base of the ice shelf, we prescribe a Robin type boundary condition, where the buoyancy pressure is a function of the vertical displacement $u_z$ given by $\rho_s g \left( h_w - u_z \right)$. A free slip boundary condition is applied to the far left terminus to restrain horizontal displacement, and allow vertical displacement that might arise due to deformation. The ocean-water pressure is applied in the direction normal to the far right terminus, increasing linearly with depth. The elevation of the ocean surface from the undeformed basal surface of the glacier is calculated using the ratio between the density of ice and ocean water $h_w = \rho_i / \rho_s \approx 90\%$. The geometry is discretised by means of approximately 450,000 triangular plane strain linear elements.

\begin{figure} [H]
  \centering
    \includegraphics[width =0.80\textwidth]{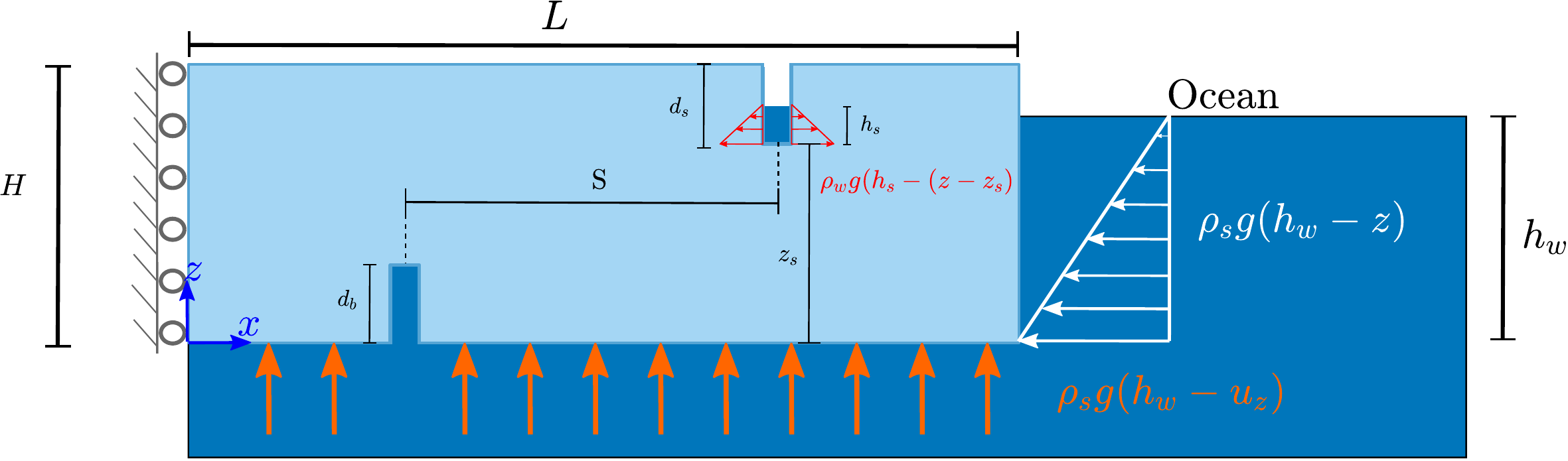}
    \caption{Growth of surface and basal crevasses in a floating ice shelf. Diagram showing the geometry and boundary conditions.}
    \label{fig:Floating BCs}
\end{figure}

\subsubsection{Propagation of surface crevasses}
\label{Sec: Floating BCs}

We first consider the finite element predictions of the longitudinal stress profile within a pristine ice shelf at different horizontal locations. Specifically, we obtain stress distributions at positions $x = \left[0, 2500, 4500, 4950\right]$ m, measured from the left edge of the glacier. The numerical predictions are then compared with the analytical solution, derived from the theory of elasticity, which is given in \ref{Appendix A}. The results are shown in Fig. \ref{fig:Stress_Distribution_Floating}, where it can be seen that the stress profiles at far field horizontal locations such as $x = 0.5$L (2500 m) are in good agreement with the predictions obtained from Eq. (\ref{eqn: Longitudinal Stress}), whereas there is a deviation from the analytical solution at locations $x = 0.95$L (4500 m) and $x = 0.99$L (4950 m), near the far right terminus. This edge effect is apparent over a greater horizontal distance when compared with the grounded glacier scenario, and is a consequence of the bending moment at the terminus due to the triangularly distributed seawater pressure. Hereafter, we only investigate the propagation of surface crevasses at horizontal locations far away from the terminus for increasing values of meltwater depth ratios and compare them with LEFM predictions based on the analytical stress solution.

\begin{figure} [H]
    \centering
    \includegraphics[width=0.6\textwidth]{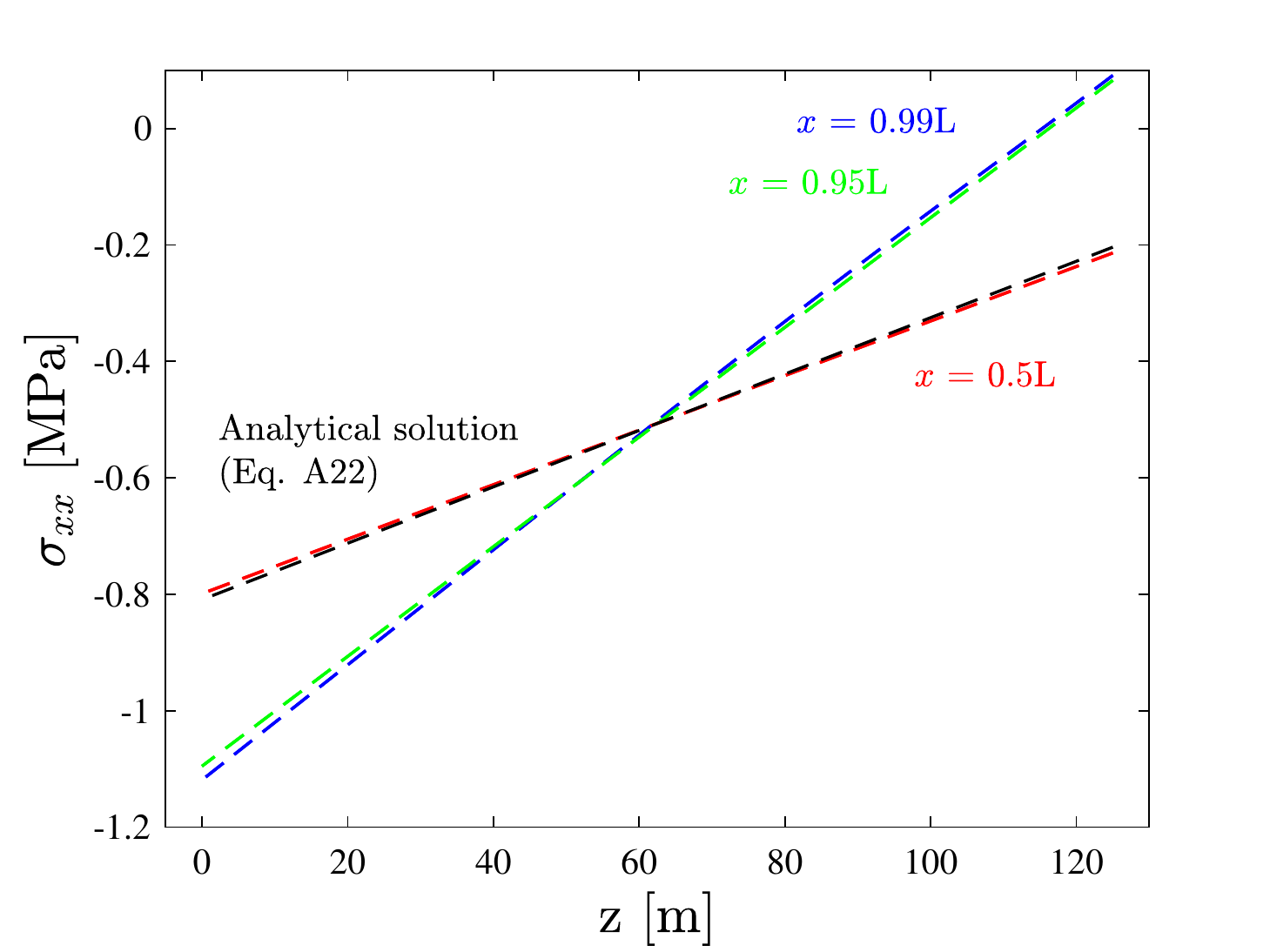}
    \caption{Surface crevasse in a floating ice shelf. Distribution of longitudinal stress $\sigma_{xx}$ versus depth at different horizontal positions. The numerical predictions are compared to the analytical solution, given in \ref{Appendix A}.}
    \label{fig:Stress_Distribution_Floating}
\end{figure}

The change in basal boundary condition from the free slip grounded condition to the Robin-type floating condition means that the double edge crack formulation is no longer appropriate for floating ice shelves. To determine the appropriate weight function for the stress intensity factor in floating ice shelves, Jim\'enez \textit{et al.} \cite{Jimenez2018OnMechanics} compared various formulations for calculating $K_{I}^{net}$ with numerically computed stress intensity factors using the displacement correlation method. It was found that the single edge crack weighting function was the most appropriate for a floating ice shelf, as given by Krug et al. \cite{Krug2014CombiningCalving}. For the different horizontal locations, we determine the appropriate relation for the longitudinal stress distribution as function of the vertical coordinate $z$ from finite element simulations and use it to evaluate stress intensity factors with Eq. (\ref{eqn: LEFM Krug}).\\

\begin{figure} [H]
    \centering
    \includegraphics[width = 0.75\textwidth]{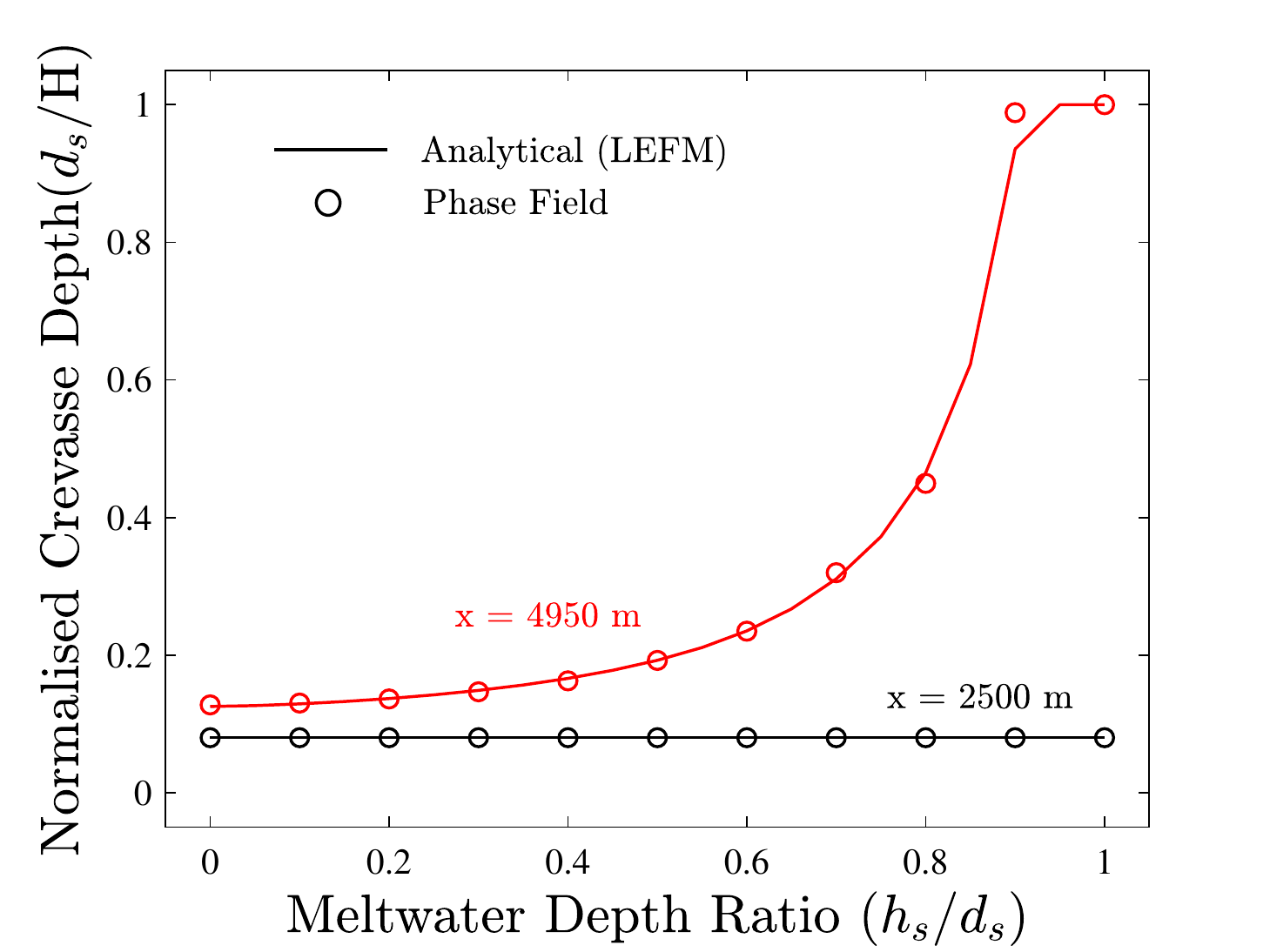}
    \caption{Growth of a surface crevasse in a floating ice shelf. Analytical (LEFM-based) and computational phase field predictions of stabilised crevasse depths as a function of the meltwater depth ratio ($h_s/d_s$). The results are provided at horizontal locations $x = 2500$ m and $x = 4950$ m.}
    \label{fig:Phase Field vs LEFM Floating}
\end{figure}

The analytical and computational predictions of stabilised surface crevasse depths within floating ice shelves are plotted in Fig. \ref{fig:Phase Field vs LEFM Floating}. For locations within the far field region (i.e. $x = 2500$ m), the longitudinal stress profile is compressive throughout the entire depth and there is no meltwater depth that will cause the crevasse to propagate beyond its initial depth of 10 m. However, surface crevasses that are close to the terminus are vulnerable to full penetration at higher meltwater depth ratios. The phase field model gives good agreement with the LEFM model for floating ice shelves when using the longitudinal stress distribution obtained from the finite element simulation and the weight function given in Eq. (\ref{eqn: LEFM Krug}).

\subsubsection{Interaction Between Surface and Basal Crevasses}

\begin{figure}[H]
\begin{subfigure}{0.5\textwidth}
  \centering
  \includegraphics[width = 0.75\textwidth]{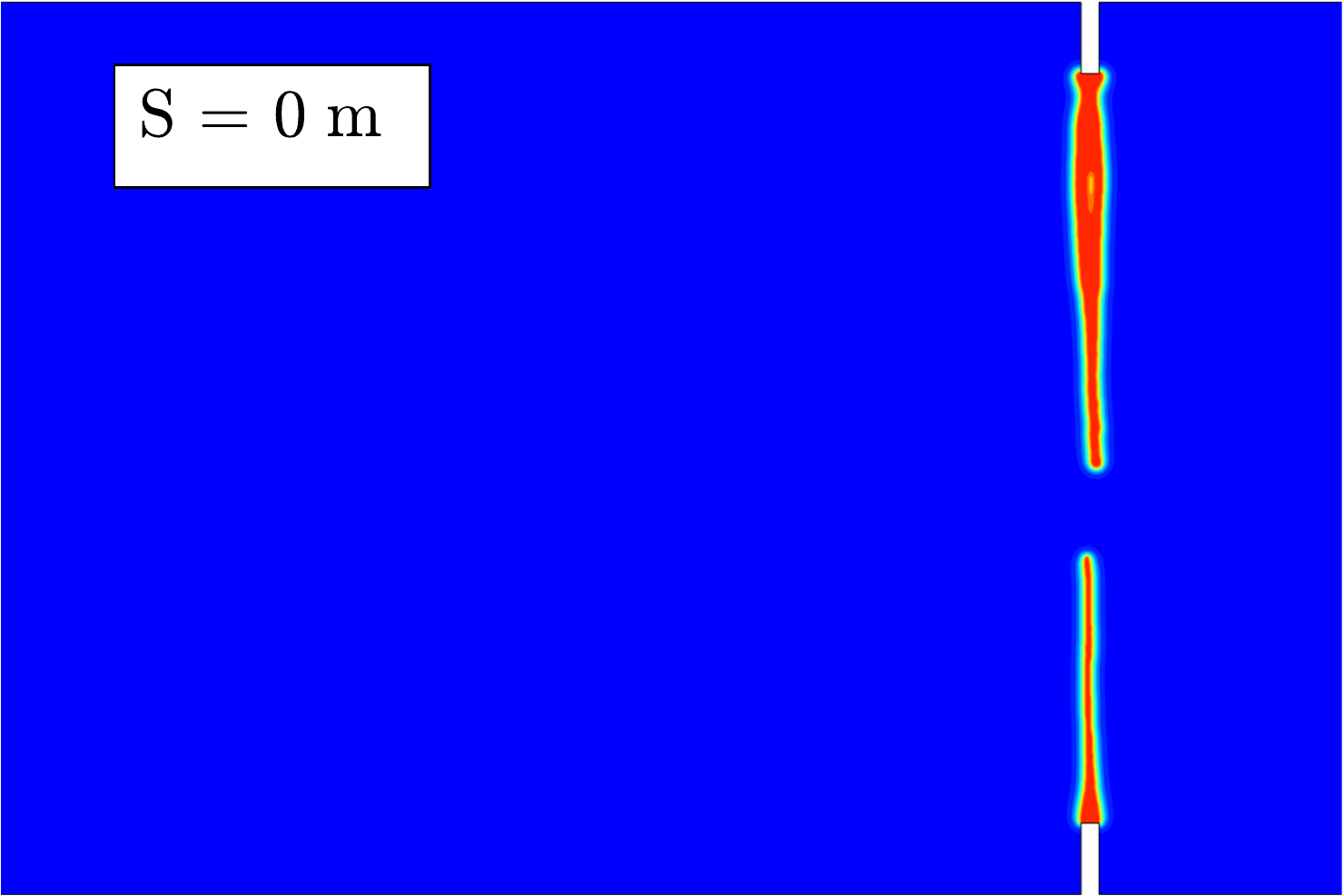}  
  \caption{}
  \label{fig:Basal Surface Directly Beneath}
\end{subfigure}
\begin{subfigure}{0.5\textwidth}
  \centering
  \includegraphics[width = 0.75\textwidth]{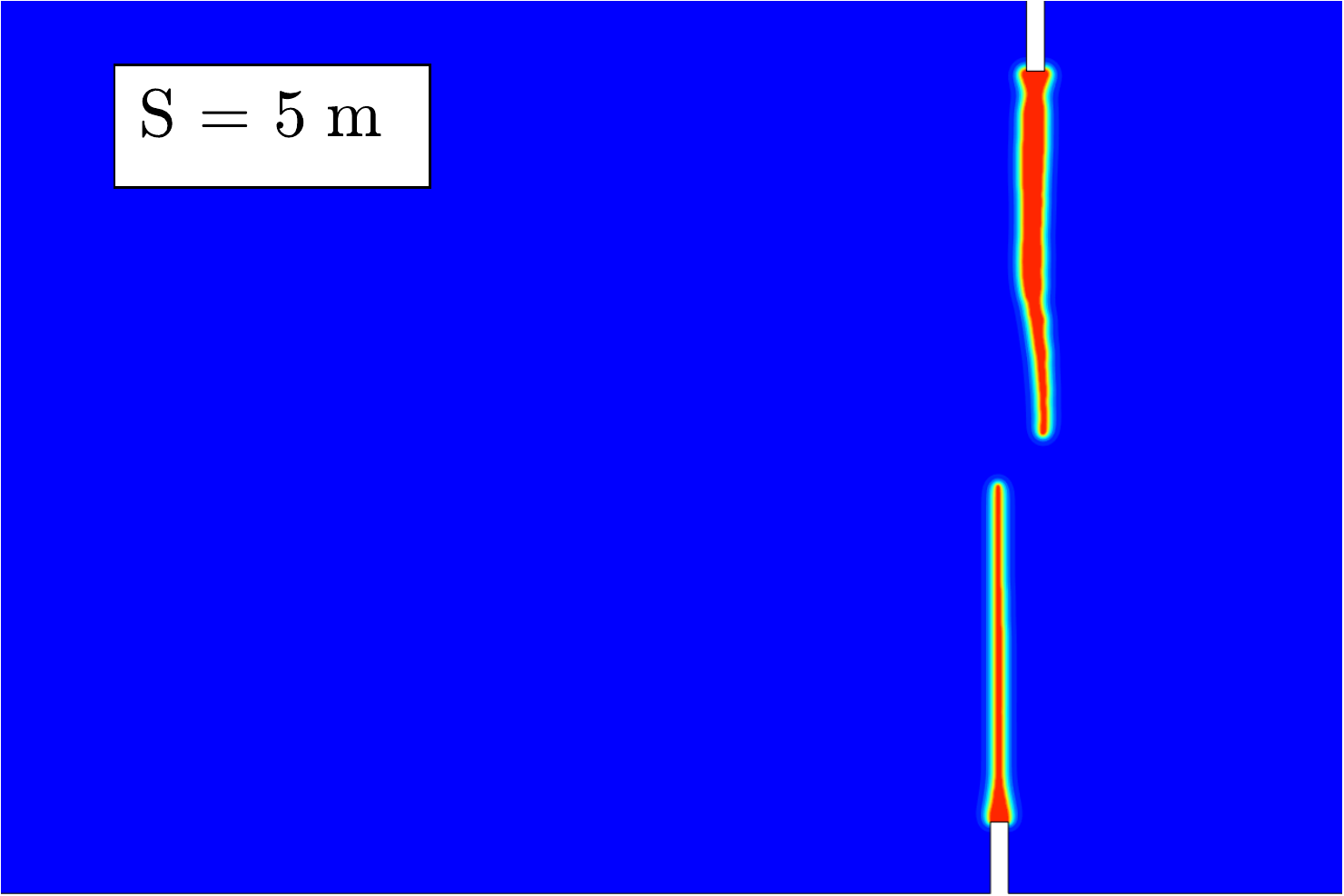} 
  \caption{}
  \label{fig:Basal Surface 5m}
\end{subfigure}
\begin{subfigure}{0.5\textwidth}
  \centering
  \includegraphics[width = 0.75\textwidth]{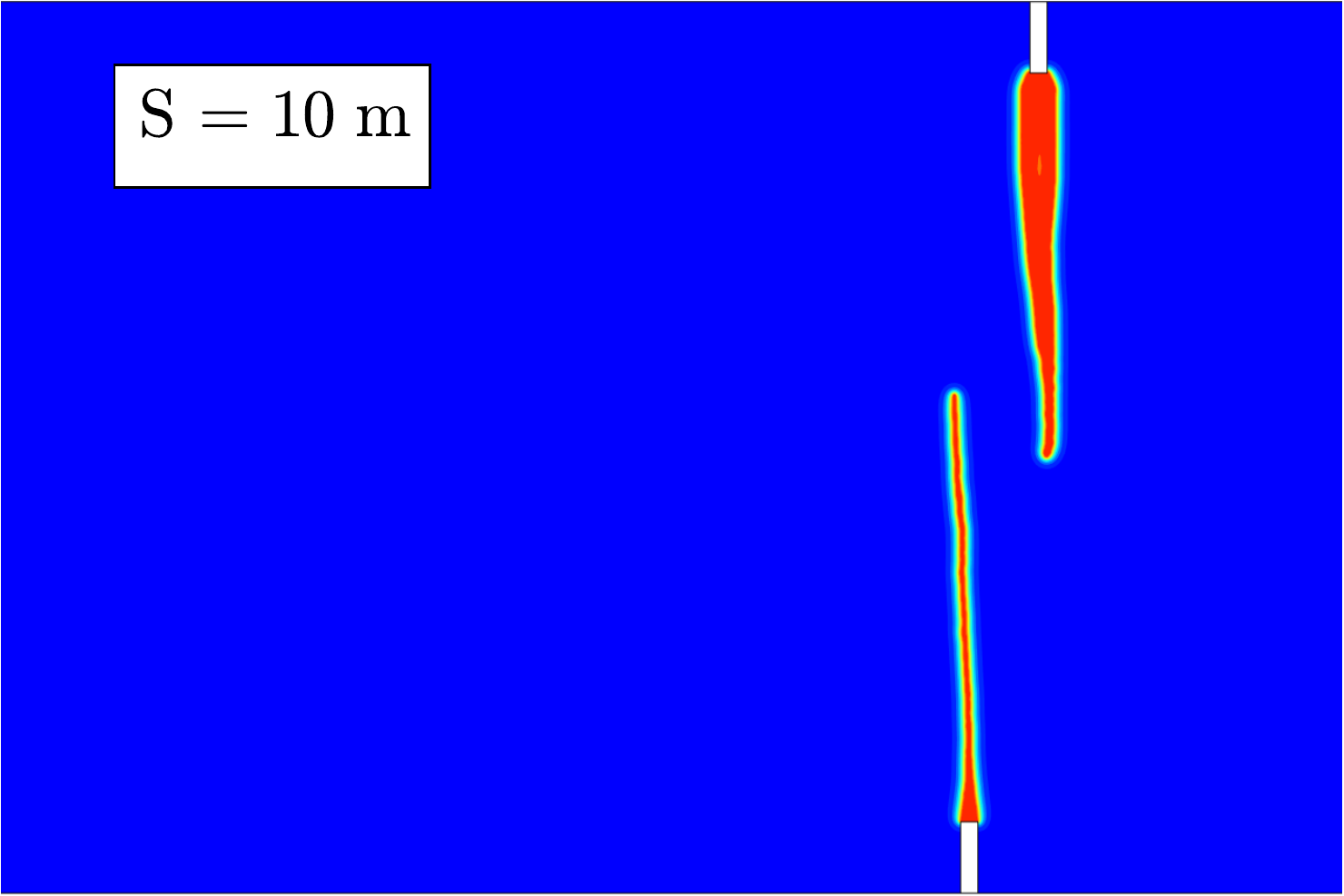} 
  \caption{}
  \label{fig:Basal Surface 10m}
\end{subfigure}
\begin{subfigure}{0.5\textwidth}
  \centering
  \includegraphics[width = 0.75\textwidth]{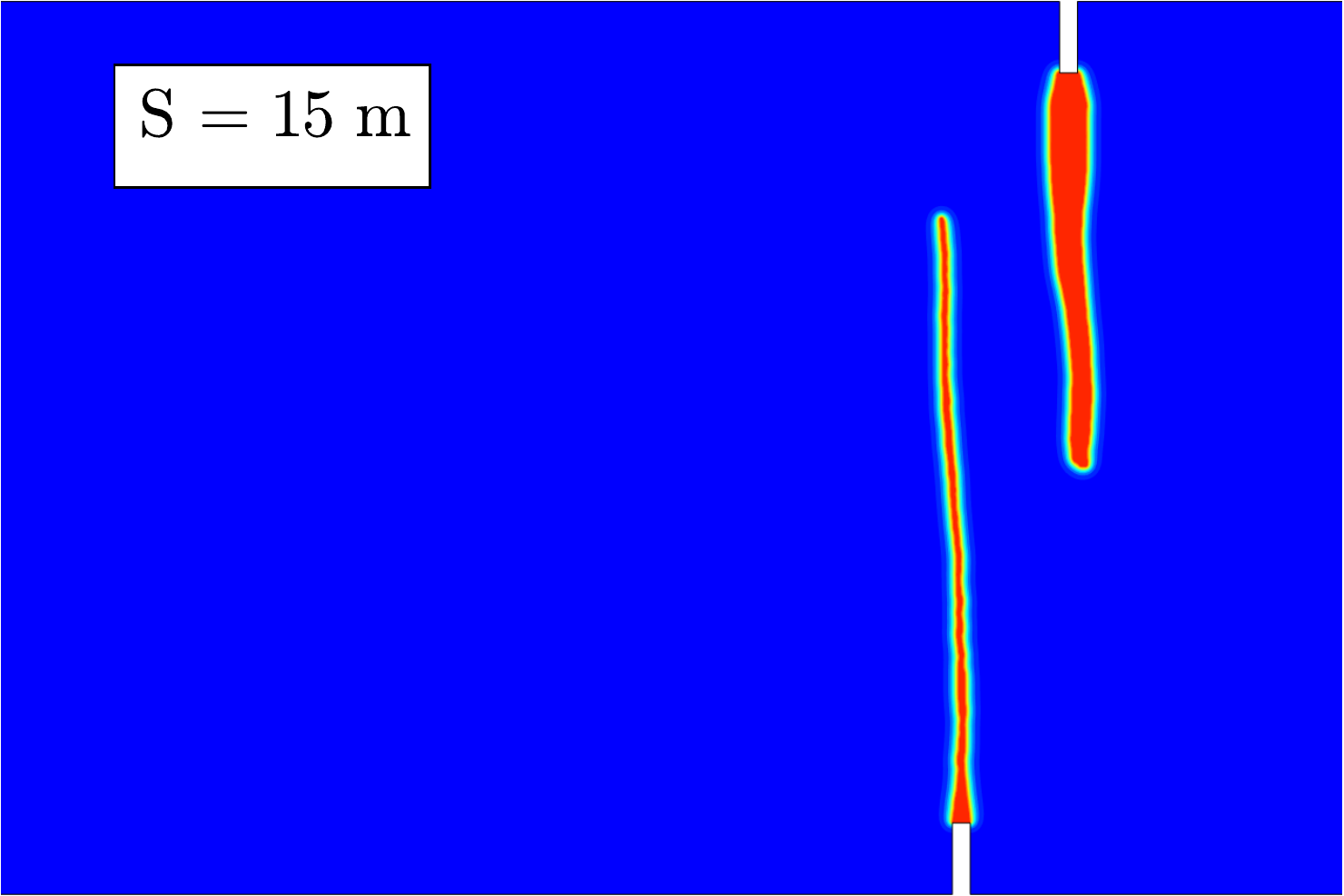} 
  \caption{}
  \label{fig:Basal Surface 15m}
\end{subfigure}
\begin{subfigure}{\textwidth}
  \centering
  \includegraphics[width = 1.05\textwidth]{Phase_Field_Legend-eps-converted-to.pdf}
\end{subfigure}
\caption{Growth of surface and basal crevasses in a floating ice shelf. Phase field damage contours after reaching the arrest of the crevasses, considering four selected values for the horizontal separation ($S$) between the basal and surface crevasse. For the surface crevasse, the meltwater depth ratio equals $h_s/d_s = 0.8$.}
\label{fig:Basal + Surface}
\end{figure}

\begin{figure}[H]
\begin{subfigure}{.5\textwidth}
  \centering
    \includegraphics[height = 7cm]{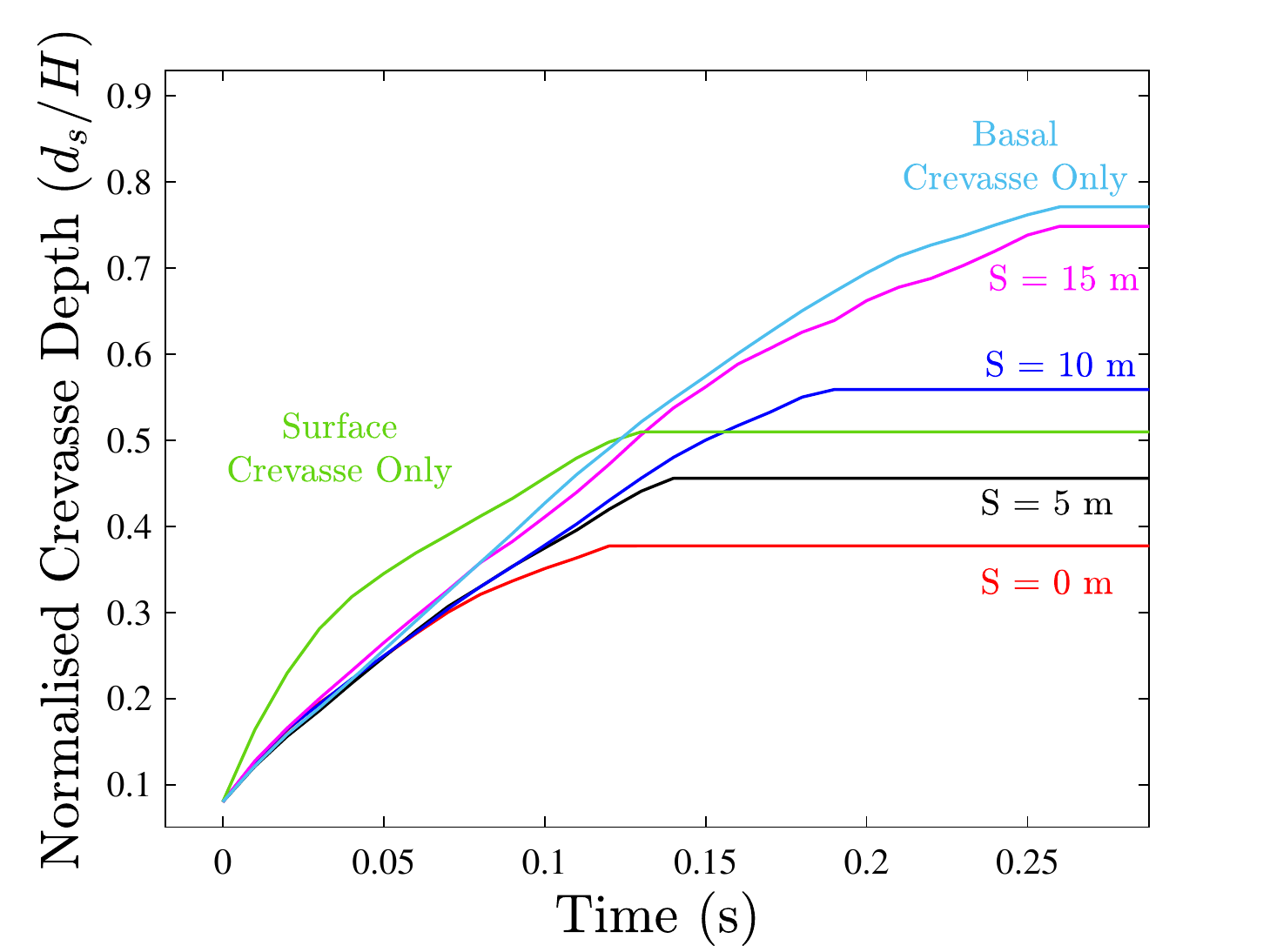}
    \caption{}
    \label{}
    \end{subfigure}
\begin{subfigure}{.5\textwidth}
  \centering
    \includegraphics[height = 7cm]{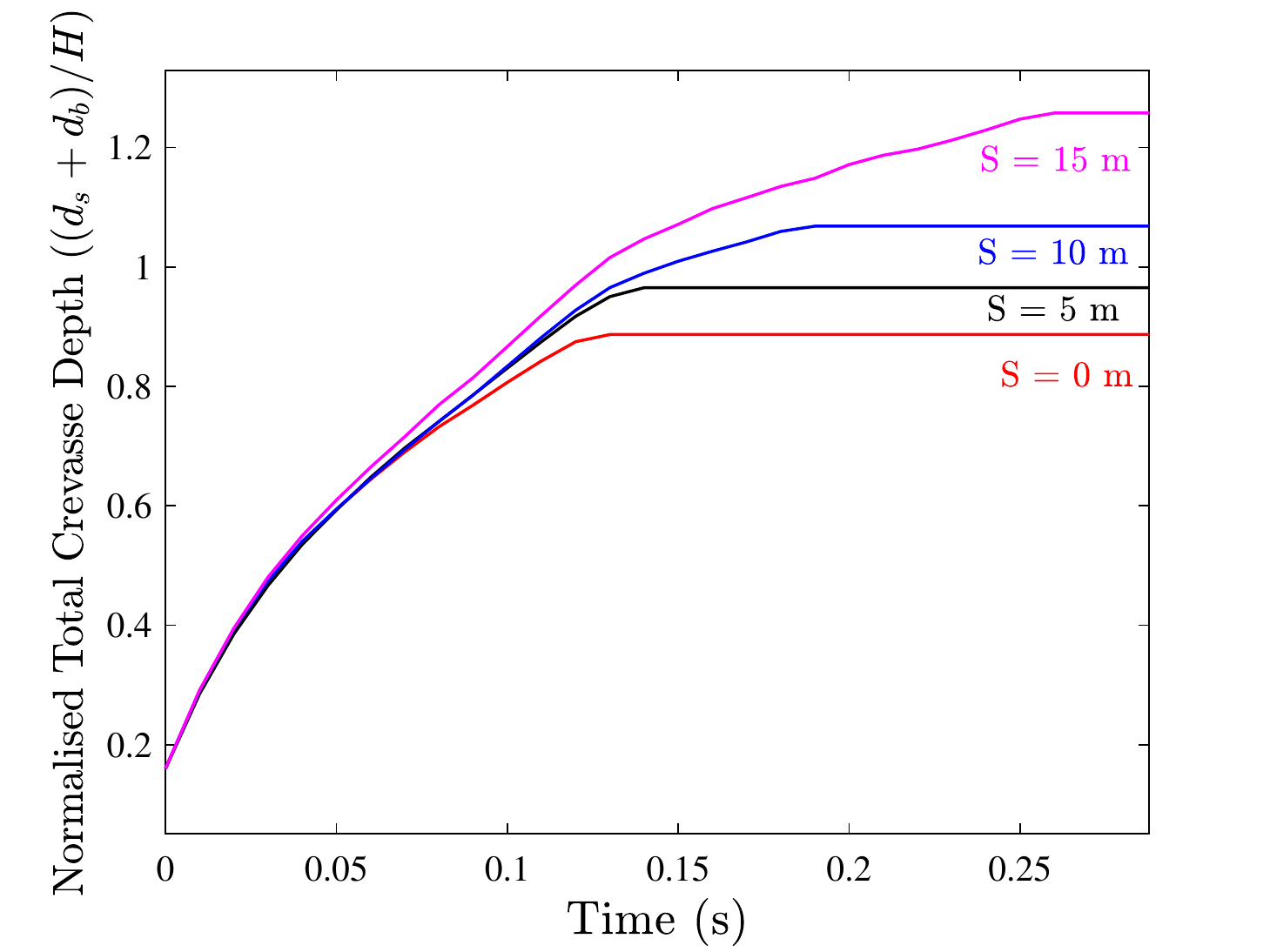}
    \caption{}
    \label{}
    \end{subfigure}
    \caption{Growth of surface and basal crevasses in a floating ice shelf. The basal and surface crevasses are separated by a horizontal distance $S$. (a) Predictions of crevasse depth versus time for surface and basal crevasses with varying horizontal spacing $S$; and (b) evolution of the combined basal and surface crevasse depth versus time for selected choices of the horizontal spacing $S$.}
    \label{fig:Basal_Surface}
\end{figure}

In a floating ice shelf, iceberg calving can occur when the combined depth of surface and basal crevasses at a location reaches the full ice thickness \cite{Nick2010ADynamics}. Therefore, we consider the propagation of a surface and a basal crevasse within close proximity of each other and near the calving front. The surface crevasse is introduced at the horizontal position $x = 4950$ m and a meltwater depth ratio of $h_s/d_s = 0.8$ is assumed, whereas the basal crevasse, located at a horizontal distance $S$ to the surface crevasse, is assumed to be fully water-filled. We consider different values of horizontal spacing $S=\{0, 5, 10, 15\}$ m between the surface and basal crevasses, to investigate if they will coalesce to form a full depth crevasse. The results obtained are shown in Figs. \ref{fig:Basal + Surface} and \ref{fig:Basal_Surface}. The phase field contours shown in Fig. \ref{fig:Basal + Surface} reveal three qualitative findings: (i) the  final depth of the surface crevasse appears to be insensitive to the presence of the basal crevasse; (ii) the depth of the basal crevasse increases with $S$, the separation to the surface crevasse; and (iii) the basal and surface crevasse do not coalescence with each other. This last effect is attributed to the mixed mode conditions that arise in the vicinity of two mode I cracks whose tips are in close proximity \cite{Nooru-Mohamed1993}.

The quantitative output of the calculations is shown in Fig. \ref{fig:Basal_Surface}. Consider first Fig. \ref{fig:Basal_Surface}a, where the predictions of crevasse depth are shown for the surface crevasse, as well as for the basal crevasse in isolation and at selected separation distances from the surface crevasse. First, a comparison with Fig. \ref{fig:Phase Field vs LEFM Floating} (for $h_s/d_s = 0.8$) shows that the extent of surface crevasse penetration is the same with and without the presence of a basal crevasse. This is unlike the basal crevasse, which exhibits a stabilised crevasse depth that it is very sensitive to the proximity of a surface crevasse. As shown in Fig. \ref{fig:Basal_Surface}a, the stabilised crevasse depth increases with the distance to the surface crevasse, with the limit case being given by the result obtained in the absence of a surface crevasse. The combined basal and surface crevasse depth is shown in Fig. \ref{fig:Basal_Surface}b. It is interesting to note that the growth of the basal crevasse is hindered by the presence of the surface crevasse when they are aligned, and consequently calving is not observed. Also, since basal and surface crevasses do not coalescence, their combined depth exceeds the glacier height for sufficiently large separations. For basal crevasses directly beneath the surface crevasse, the crevasse propagates to $37.7\%$ of the ice shelf depth, compared with $80.6\%$ for the isolated basal crevasse.

\subsection{Nucleation and growth of crevasses: application to the Helheim glacier}

In this section, we simulate the initiation and propagation of crevasses from arbitrary sites in the Helheim glacier, one of the largest outlet glaciers in southeast Greenland. The aim is to show how the creep analysis can be used to determine the nucleation of crevasses, which are then predicted to grow in a coupled deformation-fracture simulation. To generate the glacier geometry, we take the surface elevation and basal topography data from field observations (see Refs. \cite{Nick2009Large-scaleTerminus,Krug2014CombiningCalving}). A free slip boundary condition is applied normal to the base and the inlet flow velocity is restrained to zero at the left edge. Also, we apply an oceanwater pressure at the glacier terminus and assume an ocean water height of $h_w = 0.85H$. The geometry is discretised using approximately 140,000 triangular quadratic plane strain elements. 

\begin{figure} [H]
    \centering
    \includegraphics[width=\textwidth]{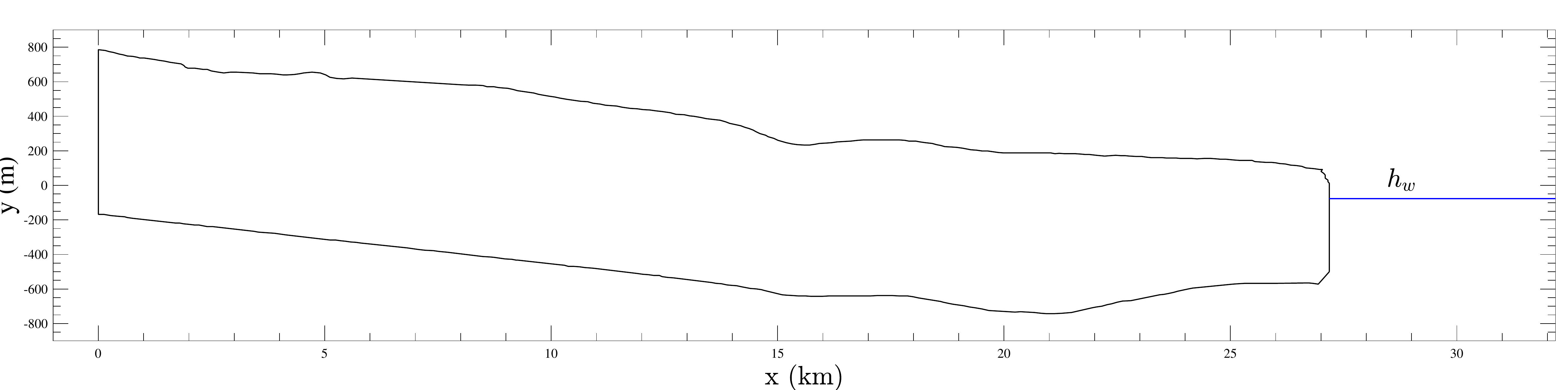}
    \caption{Nucleation and growth of crevasses in the Helheim glacier. Initial geometry, as taken from observational data by Nick \textit{et al.} \cite{Nick2009Large-scaleTerminus}.}
    \label{fig:Helheim_Geometry}
\end{figure}

\begin{figure} [H]
    \centering
    \includegraphics[width=\textwidth]{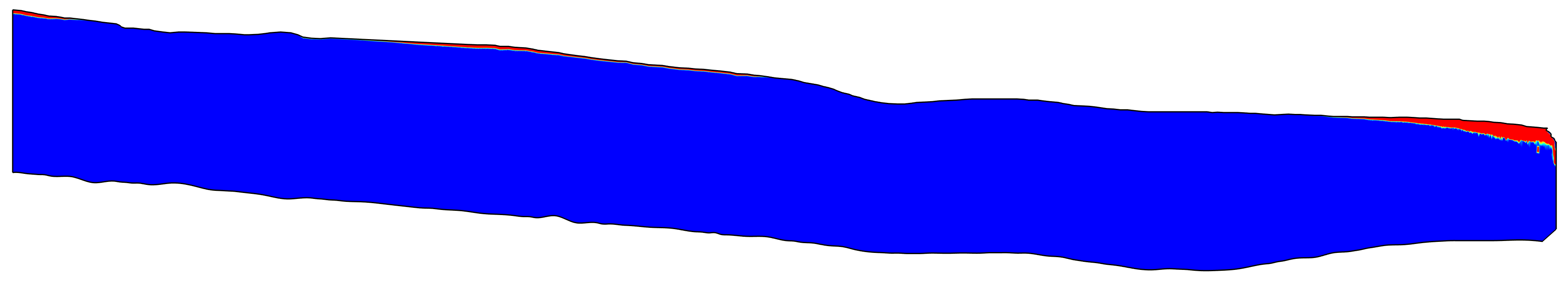}
    \caption{Nucleation and growth of crevasses in the Helheim glacier. Distribution of the nucleation variable $D_d\varepsilon^c$, with red colour contours denoting the areas where the nucleation threshold has been exceeded.}
    \label{fig:Helheim_Crack_Driving}
\end{figure}

The first step involves running a time-dependent creep simulation to determine the regions in which damage initiates. A crevasse nucleation criterion is defined by which crevasses are assumed to nucleate in regions where the product of the damage driving force state function $D_d$ and the equivalent creep strain $\varepsilon^c=\sqrt{(2/3)\bm{\varepsilon}^c:\bm{\varepsilon}^c}$ is above a certain threshold. This is denoted by red colour contours in Fig. \ref{fig:Helheim_Crack_Driving}. As it can be observed, this crack nucleation criterion is fulfilled at shallow regions within the upper surface, notably in areas with increased surface gradient and regions close to the calving front. This distribution is supported by the results by Krug \textit{et al.} \cite{Krug2014CombiningCalving}, wherein a similar pattern to initiation sites was reported from a time-dependent creep analysis. Ice is then removed from the regions where the nucleation criterion has been met, to act as initiation points for crevasse growth in the subsequent phase field step. 

\begin{figure} [H]
\begin{subfigure}{\textwidth}
  \centering
  \includegraphics[width = 0.95\textwidth]{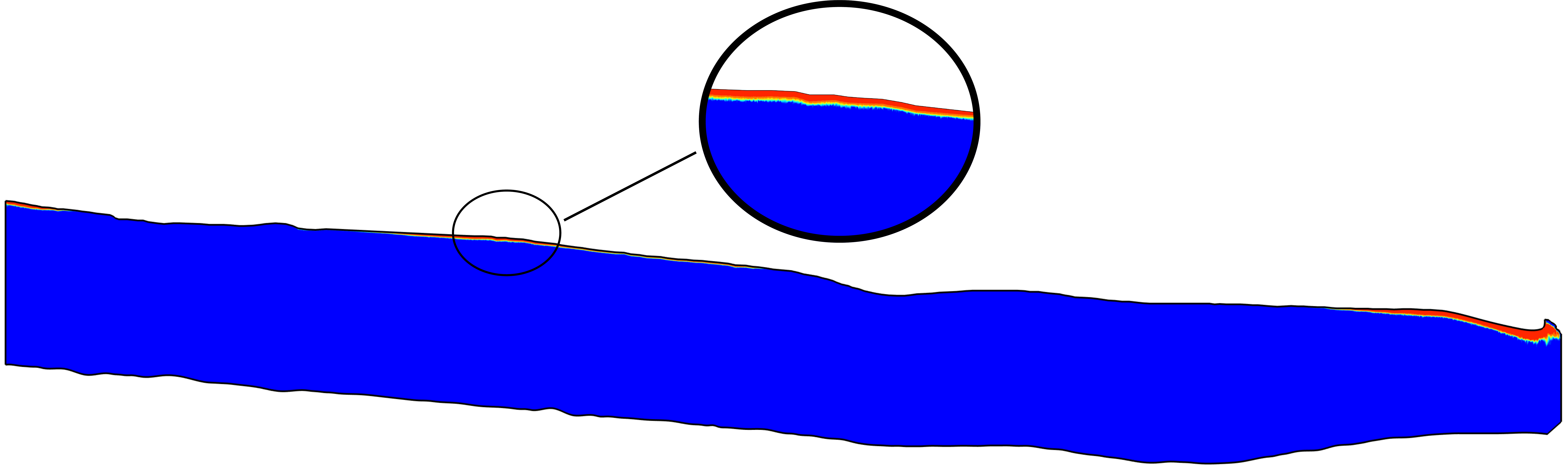}  
  \caption{}
  \label{fig:Helheim_Damage_t=0.00s}
\end{subfigure}
\begin{subfigure}{\textwidth}
  \centering
  \includegraphics[width = 0.95\textwidth]{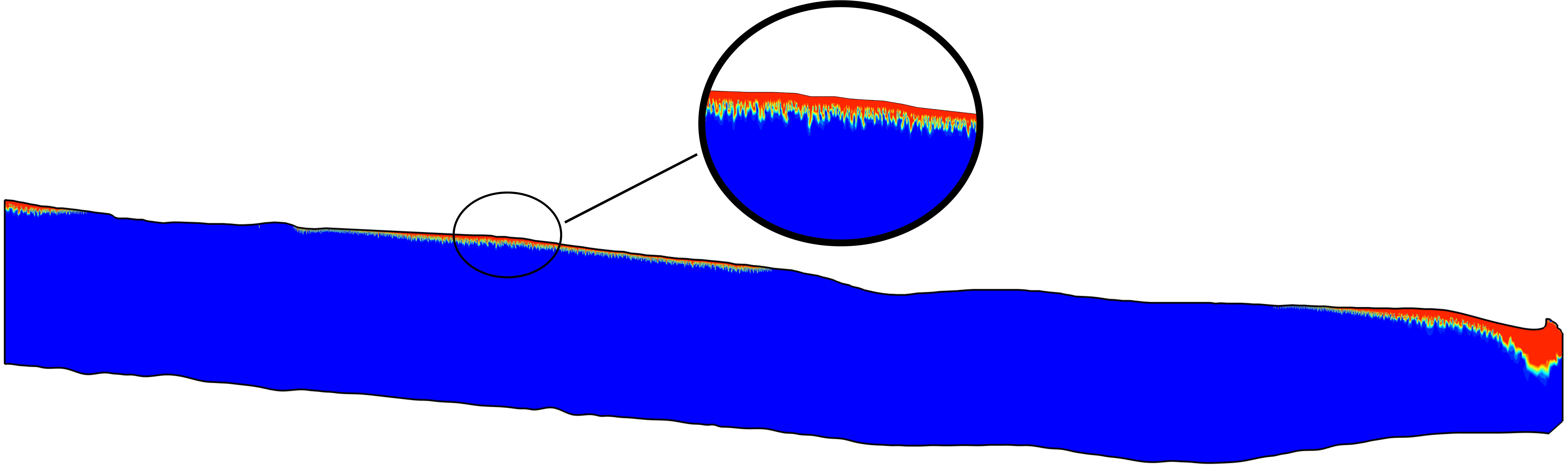}  
  \caption{}
  \label{fig:Helheim_Damage_t=0.30s}
\end{subfigure}
\begin{subfigure}{\textwidth}
  \centering
  \includegraphics[width = 0.95\textwidth]{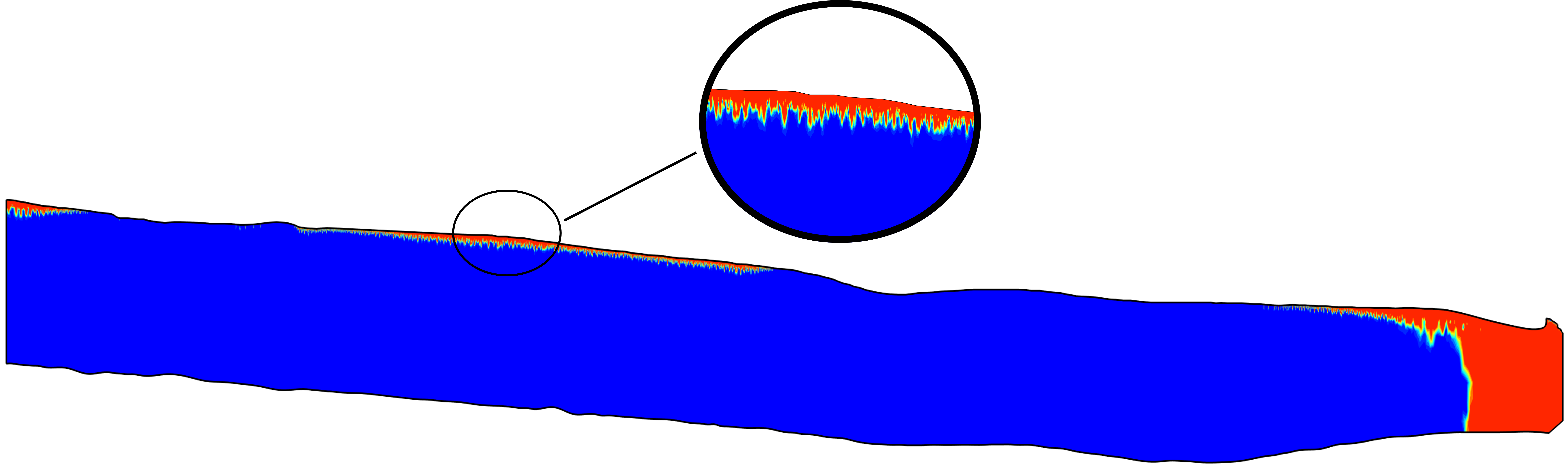}  
  \caption{}
  \label{fig:Helheim_Damage_t=1.00s}
\end{subfigure}

\begin{subfigure}{\textwidth}
  \centering
  \includegraphics[width = 1.05\textwidth]{Phase_Field_Legend-eps-converted-to.pdf}
\end{subfigure}
\caption{Nucleation and growth of crevasses in the Helheim glacier. Phase field damage evolution of the Helheim glacier assuming a non-linear viscous rheology at times (a) $t=0$ s, (b) $t=0.30$ s, and (c) $t = 0.80$ s.}
\label{fig:Helheim_Damage}
\end{figure}

Damage evolution is subsequently predicted using the phase field model with the updated geometry, assuming non-linear viscous ice rheology. As shown in Fig. \ref{fig:Helheim_Damage_t=1.00s}, we find that a field of densely spaced surface crevasses can initiate at sites both close to and away from the calving front. However, the depth at which they propagate to is shallow in comparison to the glacier geometry (approximately 40 m deep). This is in agreement with the field observations of Mottram and Benn \cite{Mottram2009}, who measured crevasse depths close to the calving front of Brei\dh{}amerkurj\"{o}kull in Iceland, finding that crevasses only penetrated tens of meters in depth. At the calving front, we also observe that damage can propagate to the full depth of the glacier, illustrating the possibility of ice cliff failure and retreat of the grounding line. This case study showcases the ability of the computational framework developed to combine creep and damage modelling to predict both the nucleation of crevasses and the subsequent propagation, for realistic geometries and conditions.

\subsection{Crevasse interactions in 3D marine-terminating ice sheets}

\begin{figure} [h]
    \centering
    \includegraphics[width = \textwidth]{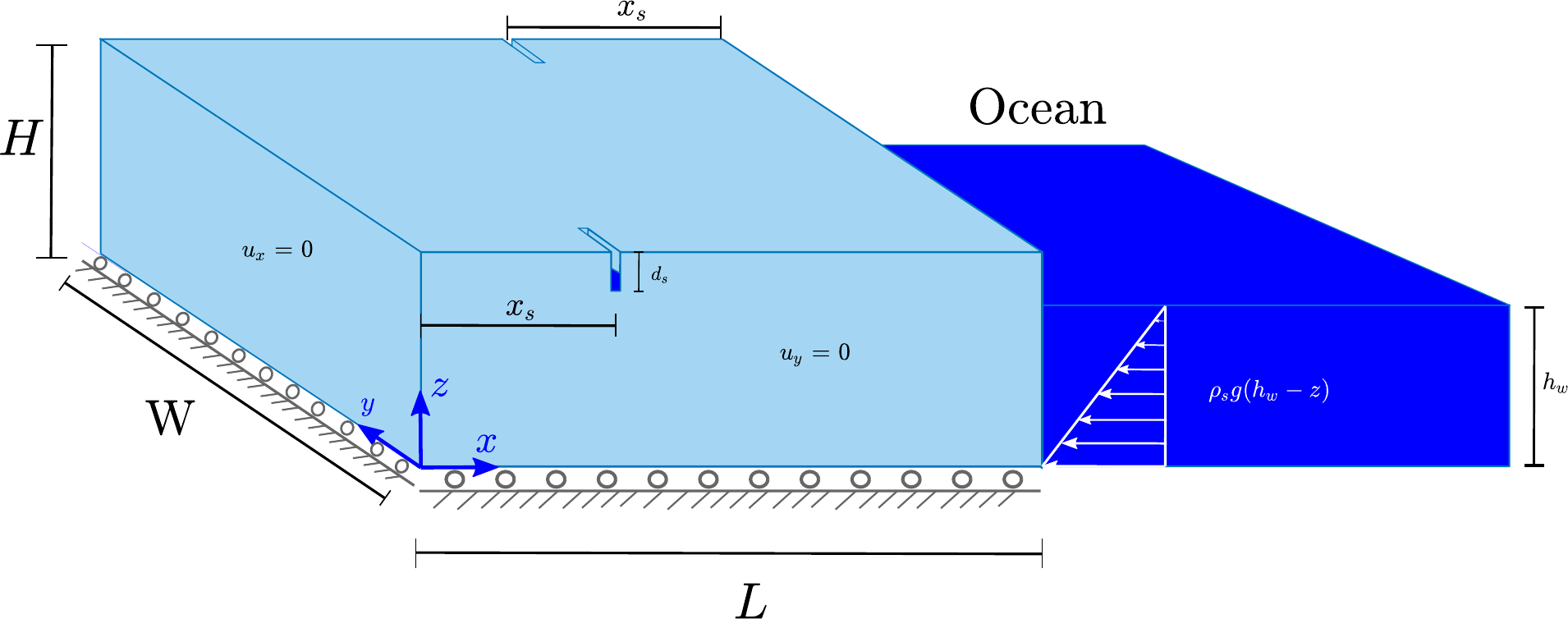}
    \caption{Crevasse interactions in 3D marine-terminating ice sheets. Diagram showing the boundary conditions and geometry of the three dimensional boundary value problem.}
    \label{fig:3D Sketch}
\end{figure}

The final numerical example intends to demonstrate the ability of the modelling framework presented to simulate damage propagation in three dimensions, including complex cracking phenomena such as crevasse interaction. We consider an idealised grounded glacier of height $H=125$ m, length $H=500$ m, and width $W=750$ m. Two dry surface crevasses are initially defined, each positioned at opposite ends of the glacier. Each crevasse is offset 25 metres either side of the centre-line in the $x$-direction, as shown in Fig. \ref{fig:3D Sketch}. Similar to the 2D plane strain case, we restrain the displacement normal to the surface at the far left edge and at the base. The displacement in the $y$-direction is also restrained at both lateral faces of the $x-z$ plane. The ocean-water pressure is applied at the far right terminus, assuming an ocean-water height of $h_w = 0.5H$. In this numerical experiment, the phase field length scale is chosen to be equal to $\ell = 10$ m; as discussed in Section \ref{Sec:Theory} and demonstrated in Section \ref{Sub: Sensitivity Analysis}, the present phase field formulation shows a negligible sensitivity to the choice of $\ell$. This enables simulating large-scale phenomena and present the first 3D ice sheet fracture simulations. The characteristic element size along the potential crevasse propagation region is chosen to be at least 5 times smaller than $\ell$ and the model is discretised using 1.5M linear tetrahedral elements.

\begin{figure} [H]
\makebox[\textwidth][c]{\centering
\begin{subfigure}{0.4\textwidth}
  \centering
  \includegraphics[height = 5cm]{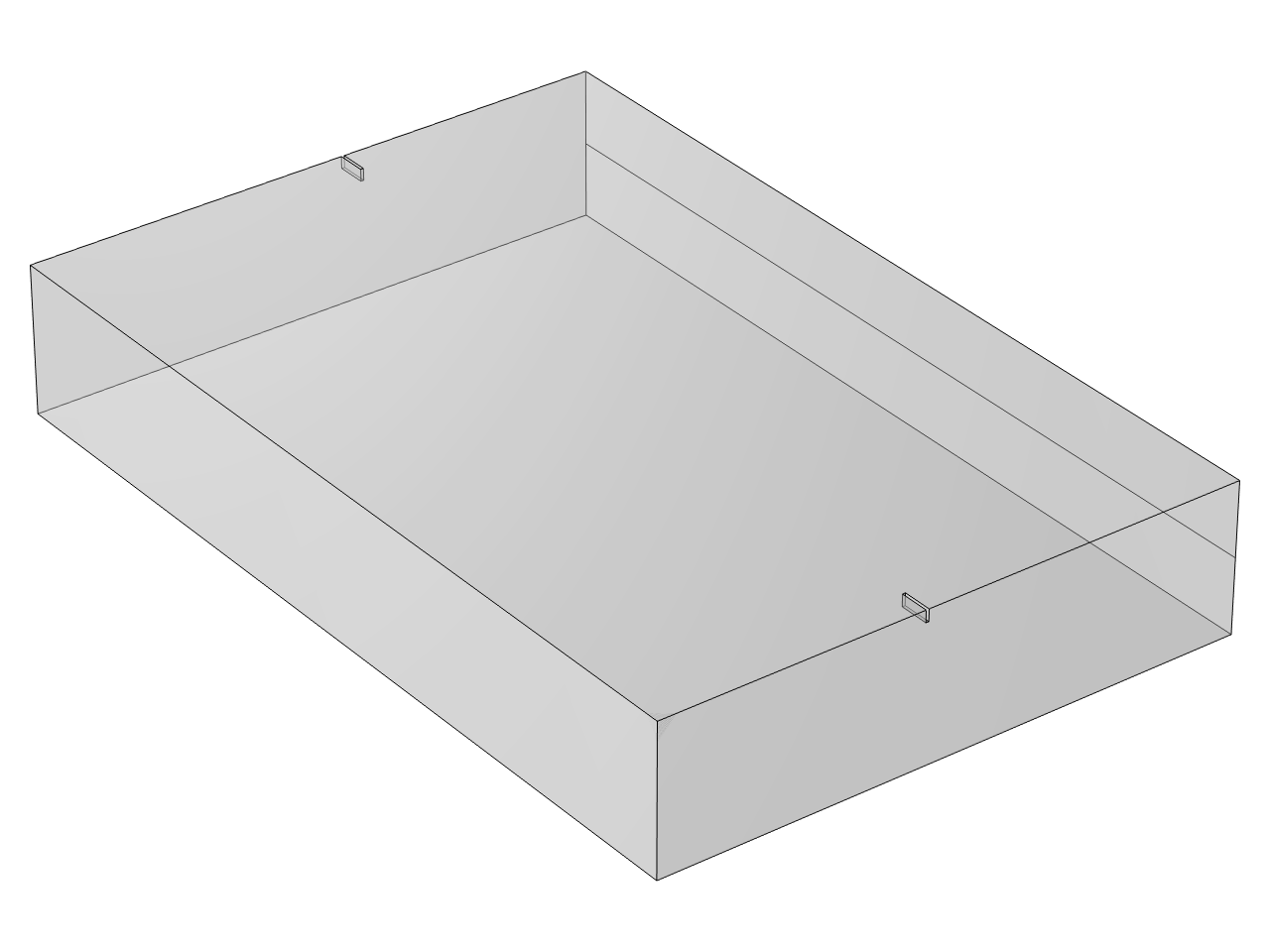}  
  \caption{}
  \label{fig:3D 0.00s}
\end{subfigure}
\begin{subfigure}{0.4\textwidth}
  \centering
  \includegraphics[height = 5cm]{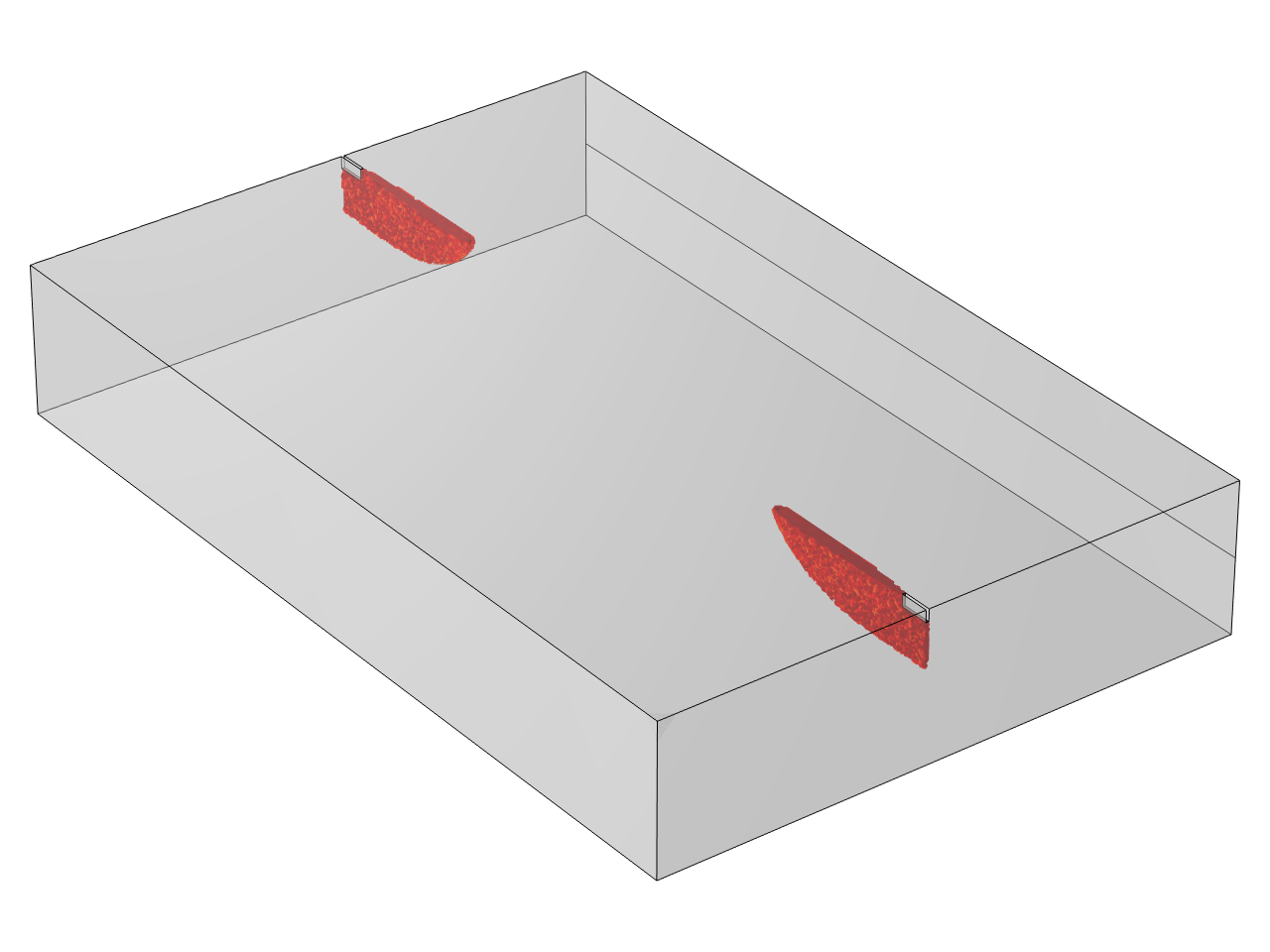} 
  \caption{}
  \label{fig:3D 0.10s}
\end{subfigure}
\begin{subfigure}{0.4\textwidth}
  \centering
  \includegraphics[height = 5cm]{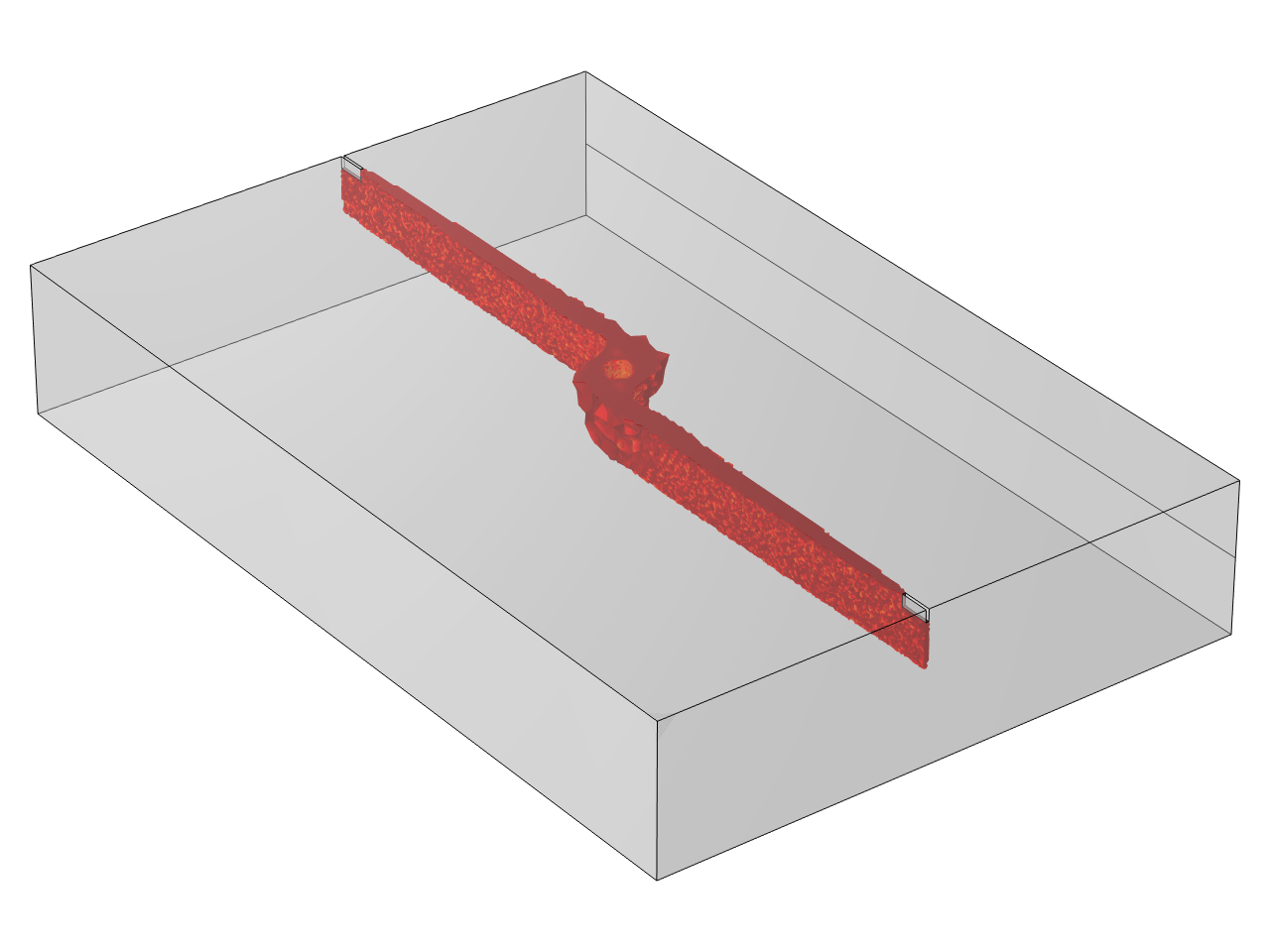} 
  \caption{}
  \label{fig:3D 0.40s}
\end{subfigure}}
\caption{Crevasse interactions in 3D ice sheets: $\phi=1$ contours showing the evolution of the two dry surface crevasses at times (a) $t = 0$ s, (b) $t = 0.05$ s, and (c) $t = 0.4$ s.}
\label{fig:3D PF}
\end{figure}

The results obtained are shown in Fig. \ref{fig:3D PF}, through contours of the phase field variable in the fully damaged regime ($\phi=1$). Initially, the two crevasses propagate vertically (along the $z$-direction) and horizontally (along the $y$-direction). Subsequently, as the two crevasses approach each other, the crack tip stress state becomes mixed mode and this results in the two crevasses curving away from each other. This is followed by the development of a hook-shaped geometry before their coalescence. Similar fracture patterns have been observed in geological faults, with remote sections of the fault growing as purely tensile fractures, whilst in close proximity to each other the faults grow as mixed mode fractures \cite{WesleyPatterson2010SegmentedBands, Acocella2000InteractionIceland}. This behavior has also been observed in laboratory experiments \cite{Nooru-Mohamed1993}.

\section{Conclusions}
\label{Sec:Concluding remarks}

We have presented a new stress-based poro-damage phase field model for predicting hydrofractures in creeping glaciers and ice shelves. The proposed framework enables resolving the underlying physical processes behind crevasse growth and iceberg calving, without the limitations and uncertainties intrinsic to widely-used empirical and analytical approaches. The model combines: (i) Glen's flow law, to adequately capture the non-linear viscous rheology of glacier ice; (ii) a poro-damage scheme that incorporates the role of meltwater pressure in assisting crevasse propagation; and (iii) a stress-based phase field description of the intact ice-crack interface. This last element is of particular importance when modelling propagating crevasses as strain energy-based phase field formulations are limited when dealing with incompressible solids and cracks driven solely by tensile stresses. The coupled framework is numerically implemented using the finite element method and used to simulate five boundary value problems of particular interest. First, the influence of the choice of material rheology and relevant parameters are investigated by simulating the propagation of a single crevasse in grounded glaciers. Second, crevasse interaction is assessed by predicting the growth of a field of densely spaced crevasses in a grounded glacier. The third case study addresses the interaction between surface and basal crevasses in a floating ice shelf, appropriately simulated using Robin boundary conditions. Nucleation and growth of crevasses in a realistic geometry, that of the Helheim glacier, is predicted in the fourth case study, combining a sequential creep-damage analysis. Finally, the last case study provides the first simulation of interacting crevasses in 3D ice sheets. Several conclusions can be obtained from the model's insight into these case studies:

\begin{itemize}
 \item The model adequately predicts the propagation of crevasses in regions where the net longitudinal stress is tensile, without the need for \textit{ad hoc} fracture driving force decompositions and exhibiting very little sensitivity to the choice of phase field length scale $\ell$. 
 
 \item Model predictions provide a good agreement with LEFM and Nye's zero stress model when particularised to the idealised conditions where these analytical approaches are relevant.
 
 \item Increasing amounts of meltwater, as a result of climate change, can significantly enhance crevasse propagation, with iceberg calving being predicted for meltwater depth ratios of 50\% or larger.
 
 \item Predicted crevasse depths are greater when considering the incompressible stress state intrinsic to a non-linear viscous rheology. Thus, first-order estimates obtained from analytical LEFM approaches should consider a Poisson's ratio of $\nu=0.5$ to avoid underpredicting the impact of meltwater on ice-sheet stability.
 
 \item The model captures how the presence of neighbouring surface crevasses provides a shielding effect on the stress concentration and reduces the predicted crevasse depth.
 
 \item The model accurately predicts the growth of surface crevasses within floating ice shelves near the shelf front for large meltwater depth ratios. Also, if a surface crevasse is in close proximity to a basal crevasse then a reduction in basal crevasse penetration depth is observed.
 
 \item Crevasses are predicted to nucleate in areas with high surface gradients, highlighting the need for an adequate characterisation of the glacier's geometry.
 
 \item The large-scale 3D analyses conducted demonstrate the capabilities of the model of opening new horizons in the modelling of crevasse growth phenomena under the computationally-demanding conditions relevant to iceberg calving.
 
\end{itemize}

Potential future extensions of the present computational framework include incorporating basal melting, lateral and basal friction effects, and ice refreezing. We offer this novel approach as a means to capture the process of crevassing and calving within ice sheets and ice shelves, to better capture these processes in efforts to prognostically assess ice-sheet vulnerability to ice shelf stability and the resulting accelerated ice sheet flow to the ocean and sea level rise, and/or grounding line retreat (potentially driven by calving at a marine-terminating margin).

\section{Acknowledgments}
\label{Acknowledge of funding}

T. Clayton acknowledges financial support from the Natural Environment Research Council (NERC) via Grantham Institute - Climate Change and the Environment (project reference 2446853). R. Duddu gratefully acknowledges the funding support provided by the National Science Foundation's Office of Polar Programs via CAREER grant no. PLR-1847173, and NASA Cryosphere award no. 80NSSC21K1003. E. Mart\'{\i}nez-Pa\~neda acknowledges financial support from UKRI's Future Leaders Fellowship programme [grant MR/V024124/1].


\appendix
\section{Derivation of the far field longitudinal stress} 
\label{Appendix A}
In this appendix the derivation of the far field longitudinal stress $\sigma_{xx}$ is presented for the grounded glacier through the equilibrium equations and Hooke's law of linear elasticity in three dimensions. The equilibrium equations for each of the three directions are as follows:
\begin{gather}
    \frac{\partial \sigma_{xx}}{\partial x} + \frac{\partial \sigma_{xy}}{\partial y} + \frac{\partial \sigma_{xz}}{\partial z} = 0 \\
    \frac{\partial \sigma_{xy}}{\partial x} + \frac{\partial \sigma_{yy}}{\partial y} + \frac{\partial \sigma_{yz}}{\partial z} = 0 \\
    \frac{\partial \sigma_{xz}}{\partial x} + \frac{\partial \sigma_{yz}}{\partial y} + \frac{\partial \sigma_{zz}}{\partial z} +\rho_i g= 0 
\end{gather}

With the assumptions of plane strain, the stresses being invariant with the x-coordinate and out of plane stresses being zero, these equations are reduced to:
\begin{gather}
     \frac{\partial \sigma_{xy}}{\partial y}  = 0 \\
     \frac{\partial \sigma_{yy}}{\partial y}  = 0 \\
     \frac{\partial \sigma_{zz}}{\partial z} +\rho_i g= 0  \label{eqn:5.6}
\end{gather}

Rearranging the equilibrium equation in the $z$-direction and integrating with respect to the vertical coordinate $z$ the vertical stress due to the lithostatic force is given by
\begin{equation}
    \sigma_{zz} = \int-\rho_i g \delta z 
\end{equation}

Substituting in the following boundary conditions 
\begin{align*}
     \sigma_{zz} & = -\rho_i g H \, \,\,\,\,\, \text{at} \,\,\,\,\, z=0 \\ 
    \sigma_{zz} & = 0  \, \,\,\,\,\, \text{at} \,\,\,\,\, z=H
\end{align*}
leads to the hydrostatic assumption of vertical stress 
\begin{equation}
         \sigma_{zz} = -\rho_i g (H - z) 
\end{equation}

The equations of linear elasticity are then used (along with the plane strain assumption) to find the out of plane normal stress $\sigma_{yy}$ in relation to the in-plane normal stresses $\sigma_{xx}$ and $\sigma_{zz}$:
\begin{gather}
    \varepsilon_{xx} = \frac{1}{E} [\sigma_{xx} - \nu(\sigma_{yy} + \sigma_{zz})]  \label{eqn:5.11} \\ 
    \varepsilon_{yy} = \frac{1}{E} [\sigma_{yy} - \nu(\sigma_{xx} + \sigma_{zz})] \\
    \varepsilon_{zz} = \frac{1}{E} [\sigma_{xx} - \nu(\sigma_{xx} + \sigma_{yy})]
\end{gather}

Setting $\varepsilon_{yy} = 0$ and rearranging to find $\sigma_{yy}$ gives:
\begin{equation}
    \sigma_{yy} = \nu (\sigma_{xx} + \sigma_{zz})
\end{equation}

Substituting this into the into the longitudinal strain equation gives:

\begin{equation}
    \varepsilon_{xx} = \frac{1}{E} \left[ (1-\nu^2) \sigma_{xx} - \nu(1+\nu) \sigma_{zz} \right] \label{eqn: x strain}
\end{equation}

The membrane stress assumption is then adopted due to the thickness of glaciers being several orders of magnitude smaller than the length. The horizontal displacement is therefore invariant with depth, leading to the the following derivative:
\begin{equation}
    \frac{\partial \varepsilon_{xx}}{\partial z} = 0
\end{equation}

Applying this constraint to Eq. \ref{eqn: x strain} and rearranging in terms of the derivative of horizontal stress gives:
\begin{equation}
    \frac{\partial \sigma_{xx}}{\partial z} = \frac{\nu}{1-\nu}\frac{\partial \sigma_{zz}}{\partial z}
\end{equation}

Substituting the above equation in Eq. \ref{eqn:5.6} yields
\begin{equation}
    \frac{\partial \sigma_{xx}}{\partial z} = -\frac{\nu}{1-\nu}\rho_i g
\end{equation}

Since the longitudinal stress is invariant with $x$-coordinate and with the plane strain assumption, the longitudinal stress is only variant on the $z$-coordinate.
\begin{equation}
    \sigma_{xx} =  - \frac{\nu}{1-\nu} \rho_i g z + C_1
\end{equation}
where $C_1$ is the indefinite integration constant that can be determined by considering the force equilibrium in the longitudinal direction for the lithostatic force of ice and the hydrostatic force of the ocean water $F_w = \frac{1}{2}\rho_s g h_w^2$ as 
\begin{equation} \label{eqn: Force Balance}
    \sum F_x = \int^H_0 \sigma_{xx}dz + F_w = 0
\end{equation}

Evaluating the definite integral in Eq. (\ref{eqn: Force Balance}) allows for the constant $C_1$ to be determined as follows:
\begin{equation}
    \left[ -\frac{\nu}{1-\nu} \frac{\rho_i g z^2}{2} + C_1 z \right]^H_0 = -\frac{\nu}{1-\nu} \frac{\rho_i g H^2}{2} + C_1H
    \end{equation}
    
    \begin{equation}
    -\frac{\nu}{1-\nu} \frac{\rho_i g H^2}{2} + C_1 H = -\rho_s g \frac{h_w^2}{2}
\end{equation}

\begin{equation}
    C_1 = \frac{\nu}{2(1-\nu)}\rho_i g H - \frac{1}{2}\rho_s g \frac{h^2_w}{H}
\end{equation}

The longitudinal stress $\sigma_{xx}$ is thus given by
\begin{equation} \label{eqn: Longitudinal Stress}
    \sigma_{xx} = \frac{\nu}{1-\nu} \left[\rho_i g \left(z-\frac{H}{2} \right) \right] - \frac{1}{2}\rho_s g \frac{h^2_w}{H}
\end{equation}

Note that the above expression does not include the effects of the meltwater pressure acting within the crevasse, which creates an additional tensile stress. The meltwater pressure $p_w$ is added to $\sigma_{xx}$ to give the net longitudinal stress used in LEFM and Nye zero stress models as follows:
\begin{equation}
    p_w = \rho_wg\left<h_s-\left(z-z_s\right)\right>
\end{equation}
\begin{equation}
    \sigma_{\text{net}}(z) = \sigma_{xx}(z) + p_w(z)
\end{equation}

\section{Discussion of appropriate LEFM model for calculation of crevasse depths} \label{Appendix B}

The linear elastic fracture mechanics model considers the effect of local stress singularity by evaluating the net stress intensity factor $K_I^{\text{net}}$ at the crack tip. This is compared to the fracture toughness $K_{IC}$, which is a measure of the material's resistance to fracture. Whilst the stress intensity factor is equal to the fracture toughness, the crack will propagate in an unstable manner; however, as the crack penetrates to greater depths (where the longitudinal stress reduces) the stress intensity factor decreases and the crack will arrest when $K_I^{\text{net}}$ becomes less than  $K_{IC}$. The stress intensity factor is calculated using Eq. \ref{eqn: LEFM} and is integrated over the entire crevasse depth due to the driving stress (far field longitudinal stress) varying linearly with depth. The use of $\sigma_{\text{net}}$ allows us to incorporate the contributions of the ice self weight, the ocean-water pressure and the meltwater pressure into the stress intensity factor.  
\begin{equation} \label{eqn: LEFM}
    K_1 ^{net} = \int^d_0  M_D\left( \chi , H, d \right) \sigma_{net} \left( \chi \right) d\chi.
\end{equation}

We evaluate the stress intensity factor using an iterative code in MATLAB by gradually increasing the crevasse depth to find the vertical coordinate where the arrest condition is met. Typical values of $K_{IC}$ for glacial ice have been determined from experimental data and are in the range of $K_{IC} = 0.1 - 0.4$ $\text{MPa}\sqrt{\text{m}}$ \cite{Rist1999, Rist2002, Fischer1995}. For this study a value of $K_{IC} = 0.1$ $\text{MPa}\sqrt{\text{m}}$ was chosen.\\

In Eq. \ref{eqn: LEFM}, $M_D$ is a weight function that is dependent upon the boundary conditions and specimen geometry. Owing to the boundary condition differences, the appropriate weight functions for the grounded glacier and the floating condition cases are different. For the grounded glacier condition, the 'double edge cracks' formulation gives good agreement with the stress intensity factors calculated using the displacement correlation method within FEM \cite{Jimenez2018OnMechanics}. The weight function for the double edge cracks model is given by \cite{Tada1985The1973}
\begin{equation}
    M_D = \frac{2}{\sqrt{2H}} \left[ 1+ f_1 \left( \frac{\chi}{d} \right) f_2 \left( \frac{d}{H} \right)   \right] \theta \left( \frac{d}{H}, \frac{\chi}{H} \right),
\end{equation}
where $\chi = H - d$, $d$ is the trial crevasse depth, and the functions $f_1, f_2$ and $\theta$ are defined as:
\begin{equation}
    f_1 = 0.3\left[ 1-\left( \frac{\chi}{d}\right)^{\frac{5}{4}} \right],
\end{equation}
\begin{equation}
    f_2 = \frac{1}{2} \left[ 1-\sin\left(\frac{\pi d}{2H}\right) \right] \left[ 2+\sin \left(\frac{\pi d}{2H}\right) \right],
\end{equation}
\begin{equation}
    \theta = \frac{\sqrt{\tan(\frac{\pi d}{2H})}}{\sqrt{1-\left[ \frac{\cos(\frac{\pi d}{2H})}{\cos(\frac{\pi \chi}{2H}) } \right]}}.
\end{equation}

For the floating ice shelf condition, the stress intensity factors calculated using the weight function method in Krug et al. \cite{Krug2014CombiningCalving} and van der Veen \cite{VanDerVeen1998a} give better agreement with the stress intensity factors calculated using the displacement correlation method \cite{Jimenez2018OnMechanics}. The formulation for the weight function $\beta$ used by Krug et al. is given below:
\begin{equation} \label{eqn: LEFM Krug}
    K_1 ^{net} = \int^d_0  \beta\left( z , H, d \right) \left( \sigma_{xx} + p_w(z) \right) \left( \chi \right) d\chi
\end{equation}
where 
\begin{equation}
    \beta\left( z, H, d \right) = \frac{2}{\sqrt{2 \pi  \left( d-z \right)}} \left[ 1+ M_1\left( 1- \frac{z}{d} \right)^{0.5} + M_2\left( 1- \frac{z}{d} \right) + M_3 \left( 1- \frac{z}{d} \right)^{1.5} \right],
\end{equation}
\begin{multline}
    M_1 = 0.0719768 - 1.513476\lambda - 61.1001\lambda^2 + 1554.95\lambda^3 - 14583.8\lambda^4 + 71590.7\lambda^5 \\
     - 205384\lambda^6 + 356469\lambda^7 - 368270\lambda^8 + 208233\lambda^9 - 49544\lambda^{10},
\end{multline}
\begin{multline}
   M_2 = 0.246984 + 6.47583\lambda + 176.456\lambda^2 - 4058.76\lambda^3 + 37303.8\lambda^4 - 181755\lambda^5 \\
   + 520551\lambda^6 - 904370\lambda^7 + 936863\lambda^8 - 531940\lambda^9 + 12729\lambda^{10},
\end{multline}
\begin{multline}
    M_3 = 0.529659 - 22.3235\lambda + 532.074\lambda^2 - 5479.53\lambda^3 + 28592.2\lambda^4 \\
    - 81388.6\lambda^5 + 128746\lambda^6 - 106246\lambda^7 + 35780.7\lambda^8,
\end{multline}

and $\lambda = d/H$.


\bibliographystyle{elsarticle-num}

\end{document}